\documentclass[journal,draftcls,onecolumn,12pt,twoside]{IEEEtranTCOM}

\usepackage{graphicx}
\usepackage{amsmath}

\usepackage{amsfonts}
\usepackage[caption=false,font=footnotesize]{subfig}
\usepackage{stfloats}
\usepackage[noadjust]{cite}
\usepackage{url}
\usepackage[noend]{algorithmic}
\usepackage{algorithm}
\usepackage{setspace}
\usepackage{multirow}
\begin{document}%
\title{Generic Multiuser Coordinated Beamforming for Underlay Spectrum Sharing}
\author{Daniel~Denkovski,~\IEEEmembership{Student~Member,~IEEE,}
				Valentin~Rakovic,~\IEEEmembership{Student~Member,~IEEE,}
				Vladimir~Atanasovski,~\IEEEmembership{Senior~Member,~IEEE,}
        Liljana~Gavrilovska,~\IEEEmembership{Senior~Member,~IEEE,}
        and~Petri~M\"{a}h\"{o}nen,~\IEEEmembership{Senior~Member,~IEEE}
\thanks{D. Denkovski, V. Rakovic, V. Atanasovski and L. Gavrilovska are with the Faculty of Electrical Engineering and Information Technologies, Ss. Cyril and Methodius University in Skopje, Macedonia. E-mails: \{danield, valentin, vladimir, liljana\}\texttt{@}feit.ukim.edu.mk}
\thanks{P. M\"{a}h\"{o}nen is with the Institute for Networked Systems, RWTH Aachen University, Germany. E-mail: pma\texttt{@}mobnets.rwth-aachen.de}}

\maketitle

\begin{abstract}
The beamforming techniques have been recently studied as possible enablers for underlay spectrum sharing. The existing beamforming techniques have several common limitations: they are usually system model specific, cannot operate with arbitrary number of transmit/receive antennas, and cannot serve arbitrary number of users. Moreover, the beamforming techniques for underlay spectrum sharing do not consider the interference originating from the incumbent primary system. This work extends the common underlay sharing model by incorporating the interference originating from the incumbent system into generic combined beamforming design that can be applied on interference, broadcast or multiple access channels. The paper proposes two novel multiuser beamforming algorithms for user fairness and sum rate maximization, utilizing newly derived convex optimization problems for transmit and receive beamformers calculation in a recursive optimization. Both beamforming algorithms provide efficient operation for the interference, broadcast and multiple access channels, as well as for arbitrary number of antennas and secondary users in the system. Furthermore, the paper proposes a successive transmit/receive optimization approach that reduces the computational complexity of the proposed recursive algorithms. The results show that the proposed complexity reduction significantly improves the convergence rates and can facilitate their operation in scenarios which require agile beamformers computation.
\end{abstract}

\begin{IEEEkeywords}
Multiuser MIMO, underlay sharing, generic coordinated beamforming, sum rate maximization, fairness, recursive and successive optimization, interference, broadcast, multiple access channels.
\end{IEEEkeywords}

\IEEEpeerreviewmaketitle

\section{Introduction}
\label{sec:intro}
\IEEEPARstart{T}{he} proliferation of wireless devices and ever increasing use of mobile data makes efficient use of radio spectrum a critical issue. Cognitive Radio (CR) is identified as a promising technology that can facilitate secondary (non-licensed) users to re-utilize or share the licensed bands \cite{cite1} and significantly improves the spectrum utilization. Employing multiple transmit and receive antennas at the secondary user (SU) systems can guarantee high spectrum efficiency while avoiding interfering with the primary (incumbent) users (PUs). The use of multiple antennas offers additional degrees of freedom due to the spatial dimension \cite{cite2,cite3,cite4}, which can support simultaneous and transparent operation of the SU systems with respect to the PUs (underlay sharing). Moreover, multiple antennas provide possibilities for more flexible and efficient resource allocation compared to the conventional SU systems (e.g. interweave-based SU systems). 

\indent Recent research activities in spectrum sharing have shown considerable interest in the design of practical and efficient beamforming techniques for the underlay-based SU systems \cite{cite5,cite6,cite7,cite8}. The objective of beamforming in the context of underlay spectrum sharing is to maximize the SU rate with a given transmit power budget, while keeping the harmful interference to the PU system below a predefined threshold. 

\indent The previous works on beamforming with underlay spectrum sharing generally focus on scenarios where only one SU (communication pair or data link) shares the spectrum with one or multiple PUs \cite{cite5,cite11,cite12,cite13,cite14}. More recently, the design of the beamforming techniques has been extended and applied to the multiuser SU scenario \cite{cite7,cite8,cite15,cite16,cite12,cite18,cite10,cite19,cite20,cite21,cite22,cite23,cite24,cite25}. However, there is a lack of generic and robust transmit/receive beamforming designs applicable to different system models and operating with arbitrary number of antennas and users. The existing beamforming techniques are usually designed for a specific system model, i.e. optimized to be used for the interference channel \cite{cite10,cite21} broadcast channel \cite{cite22,cite25} or the multiple access channel \cite{cite23,cite24}. Another drawback of the existing schemes is that the maximum number of SUs in the system cannot exceed the number of either transmit or receive antennas in the system \cite{cite10,cite19,cite20,cite21,cite22,cite23,cite24,cite25}. Most of literature considers system models and beamforming techniques that are developed for the broadcast channel \cite{cite7,cite8,cite15,cite16,cite12,cite18} and the MISO transmit beamforming scenarios. To the best of the authors knowledge only Liu \& Dong in \cite{cite10} and Scutari \& Palomar in \cite{cite20} have tried to address the issue of the multiuser interference channel for the MIMO-based underlay spectrum sharing. However, the proposed solutions in \cite{cite10,cite20} can only serve a specific number of SUs, which is upper bounded by the number of transmit antennas.

\indent Even in conventional multiuser MIMO beamforming research, the authors have struggled to produce generic multiuser coordinated beamforming techniques relying on convex problems for the sum rate optimization. Most of the current work focused on the zero forcing transmit beamforming and the optimal maximum ratio combining for the receiver beamforming. Employing zero forcing \cite{cite35,cite36} at the transmit receiver side imposes limitations on the number of operating users and the number of antennas in the system. Instead of zero forcing, the sum rate optimal transmit beamformers for the multiuser MIMO broadcast and medium access channel can be also solved using the generalized BC-MAC duality approach \cite{cite38}. However, the BC-MAC duality is not feasible to underlay spectrum sharing scenarios. The sum rate optimization has been also tackled for the MISO interference channel, with the optimal transmit beamformers attained via SINR feasibility and outer polyblock approximation approaches \cite{cite37}. These approaches result in high computational complexity and generally slow convergence.

\indent In addition, most of the beamforming related underlay sharing literature omits the reverse interference, i.e. the interferences caused from the PU to the CR system, in their system models and the corresponding beamforming designs \cite{cite5,cite6,cite7,cite8,cite9,cite10,cite19,cite20,cite21,cite22,cite23,cite24,cite25}. In realistic underlay sharing scenarios, the reverse interference can severely diminish the performance of the SU system and should not be neglected when performing the beamforming optimization \cite{cite11}. In order to deal with and align with the PU interference caused to the secondary system, multiple antennas are a necessity at the SU receiver sides.

\indent In this paper we address some of the limitations of the existing work. Specifically, we propose several underlay spectrum sharing techniques with respect to the beamforming optimization. In particular, the contributions of this paper are the following. First, we propose a generic beamforming design by exploiting the advantages of the coordinated beamforming that can be utilized for underlay spectrum sharing. Our generic beamforming design can be applied for all three major system models, i.e. interference, broadcast and multiple access channels. It allows underlay operation of the arbitrary number of users and antennas. Furthermore, the proposed beamforming design can be applied to conventional MIMO beamforming systems. Second, we propose two novel beamforming algorithms that provide convex solutions for recursive computation of the optimal transmit and receive beamformers. The former leverages fairness between the users in the system, while the later one maximizes the sum rate in the system. In addition, the paper proposes a successive optimization approach that decreases the computational complexity of the presented fairness and sum rate maximization beamforming algorithms. These algorithms are also applicable to conventional MIMO beamforming systems. Finally, we present and employ a more generic system model for the underlay spectrum sharing that takes into account the reverse interference caused by the PU system to the SU system.
\begin{figure*}[!ht]
\centering
\def\imagetop#1{\vtop{\null\hbox{#1}}}
\begin{tabular}[t]{lcrr}
\subfloat[]{\includegraphics[height=1.9in, width=1.75in]{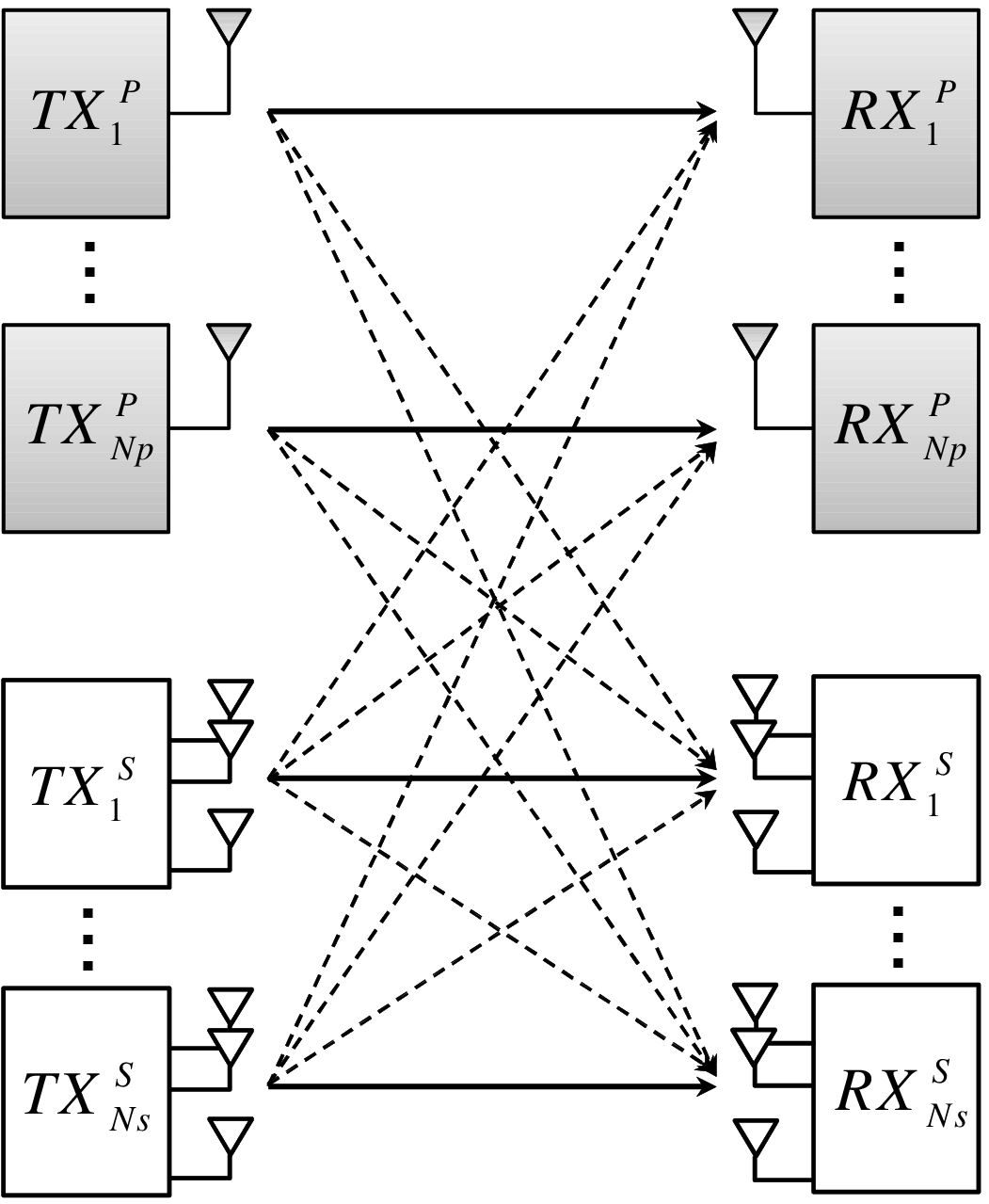}\label{fig:1a}} &
\subfloat[]{\includegraphics[height=1.9in, width=1.75in]{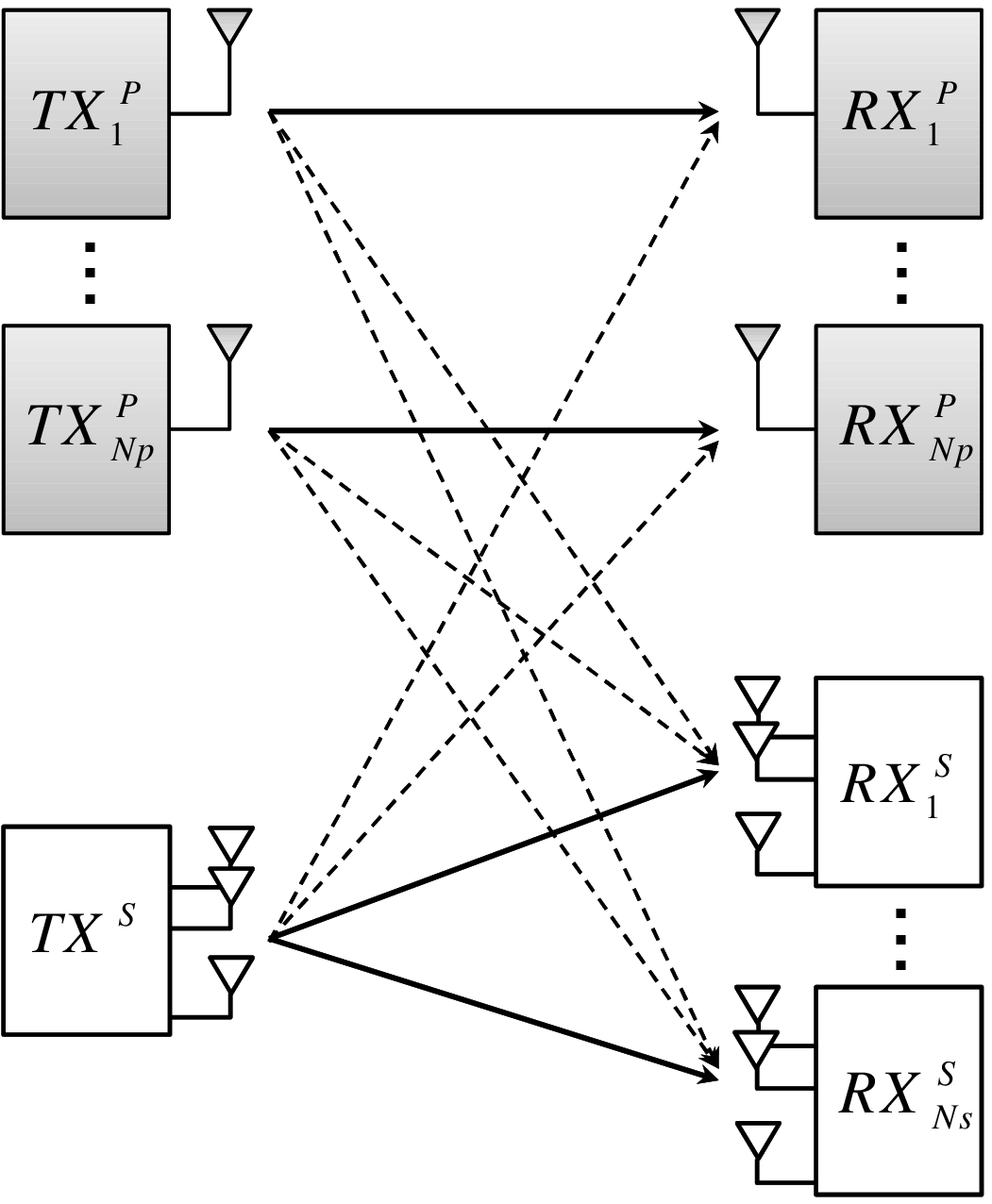}\label{fig:1b}} &
\subfloat[]{\includegraphics[height=1.9in, width=1.75in]{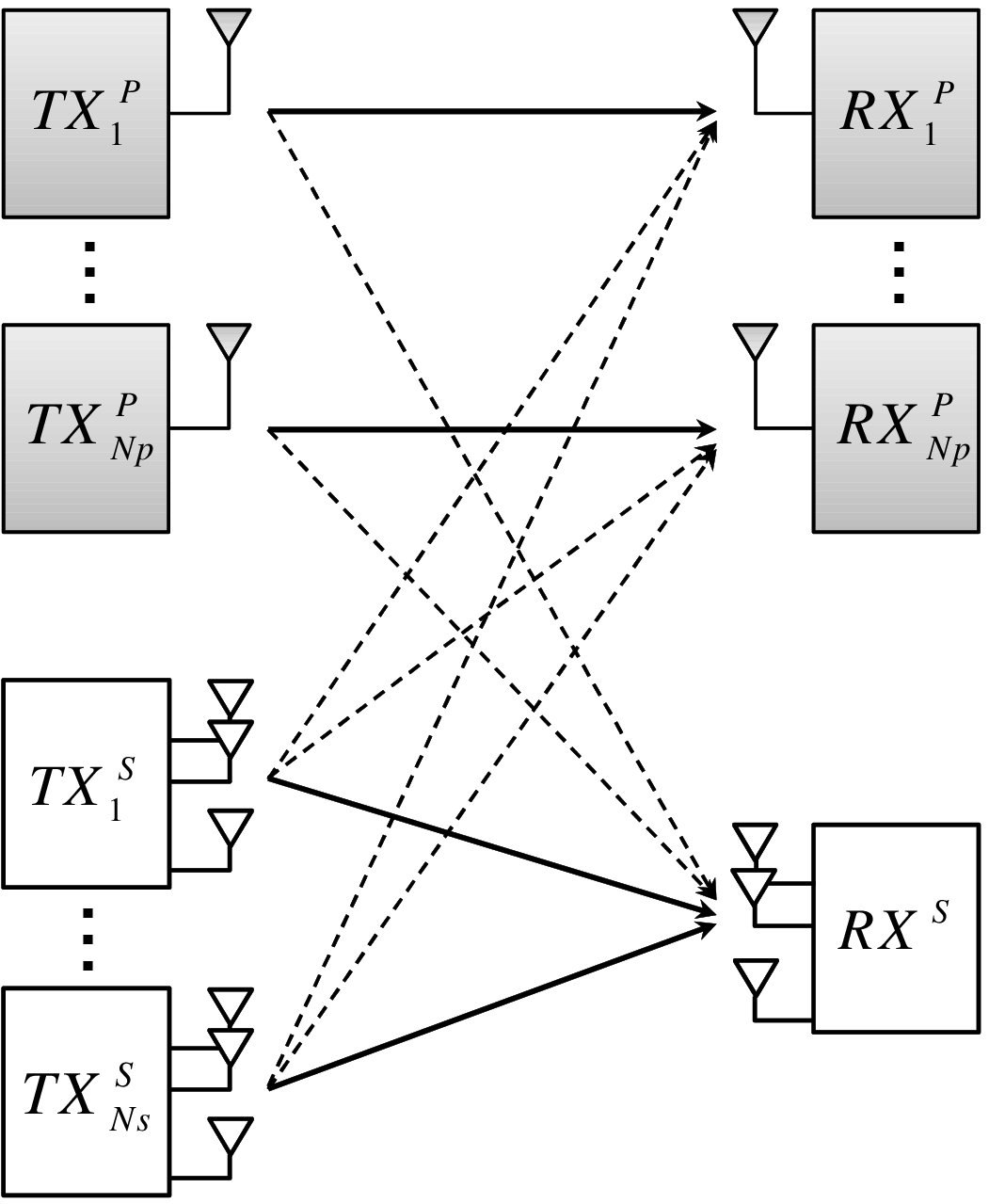}\label{fig:1c}} &
\subfloat{}{\includegraphics[height=0.5in]{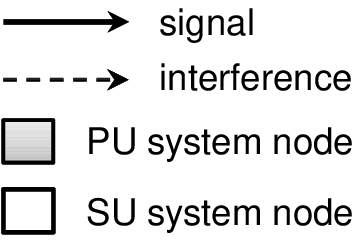}}
\end{tabular}
\caption{System model for multiuser MIMO CR (a) interference channel, (b) broadcast channel and (c) MAC channel}
\label{fig:1}
\end{figure*}

\indent The remainder of the paper is organized as follows. Section~\ref{sec:sysmodel} presents the notation and system model under consideration. Section~\ref{sec:genbf} presents the design of the recursive fairness and sum rate maximization algorithms, as well as convexity reduction for successive transmit/receive beamforming optimization. Section~\ref{sec:PerfAnl} assesses the performance of the algorithms under different system models considerations and scenario setups. Finally, Section~\ref{sec:Concl} concludes the paper.

\section{System Model}
\label{sec:sysmodel}
\indent We consider a multiuser beamforming communication of $N_s$ secondary users operating in underlay manner in the same band as $N_p$ primary user pairs. The paper considers three possible MIMO system models for the secondary user communication, i.e. a multiuser interference (IC) channel (Fig.~\ref{fig:1a}), a broadcast (BC) channel (Fig.~\ref{fig:1b}) and a multiple access (MAC) channel (Fig.~\ref{fig:1c}). The following text first considers the multiuser MIMO CR interference channel model as the most comprehensive one and afterwards presents the respective simplifications for the remaining system models. 

\indent The secondary users are equipped with $N_t$ antennas at the transmitter side and $N_r$ antennas at the receiver side, while the primary users target a SISO communication. All the channels in the system are assumed to follow a $\mathcal{CN}$ distribution. In the remaining text, $\left\|\right\|$ denotes a vector norm, $tr$ denotes a matrix trace operation, $rank$ represents a matrix rank operation and $blkdiag$ denotes a block diagonal matrix.

\subsection{Multiuser MIMO CR interference channel}
\indent In this subsection we consider the most comprehensive MIMO CR interference channel model (Fig.~\ref{fig:1a}). The secondary system is composed of $N_s$ independent communication pairs (transmitter and receiver) aiming to transparently utilize the same band used by $N_p$ communication pairs on a primary basis. The notations in the subsequent analysis use per communication pair (SU or PU) indexing. In particular, the $1\text{x}1$ channel between the transmitter in the PU pair $i$ and the receiver in the PU pair $j$ is denoted as $h_{pp(i,j)}$, where $i,j\in \left\{1,...,N_p\right\}$. The $N_r\text{x}N_t$ communication channels between the secondary users are denoted as matrices $\mathbf{H}_{ss(k,l)}$, where $k$ denotes the index of the SU pair the transmitter belongs to, $l$ denotes the index of the receiver's SU pair, and $k,l\in \left\{1,...,N_s\right\}$. The $N_r\text{x}1$ interference channel between the PU transmitter $i$ and SU receiver $l$ is represented as $\mathbf{h}_{ps(i,l)}$, where $i\in \left\{1,...,N_p\right\}$ and $l\in \left\{1,...,N_s\right\}$, while the interference channel between the SU transmitter $k$ and the PU receiver $j$ is represented with the $1\text{x}N_t$ vector $\mathbf{h}_{sp(k,j)}$, where $k\in \left\{1,...,N_s\right\}$ and $j\in \left\{1,...,N_p\right\}$. The $N_t\text{x}1$ transmit and the $N_r\text{x}1$ receive beamforming vectors for the SU pair $k$ are denoted as $\mathbf{m}_k$ and $\mathbf{w}_k$, respectively.

\subsubsection {Secondary system perspective}
\indent The signal received at the SU receiver $l$ comprises the signal from the SU transmitter $l$ summarized with the cumulative interference coming from all the PU transmitters, the cumulative interference from the remaining SU transmitters ($k\neq l$), multiplied by the respective channels, as well as the additive $\mathcal{CN}$ noise at the SU receiver $l$, i.e.
\begin{equation}\label{eq:1}
y_{s(l)}= \mathbf{w}_{l}^{H}\mathbf{H}_{ss(l,l)}\mathbf{m}_ls_{s(l)} + \sum_{i=1}^{N_p}{\mathbf{w}_{l}^{H}\mathbf{h}_{ps(i,l)}s_{p(i)}} + \sum_{k=1;k\neq l}^{N_s}{\mathbf{w}_{l}^{H}\mathbf{H}_{ss(k,l)}\mathbf{m}_ks_{s(k)}} + n_{s(l)},
\end{equation}
\noindent where $s_p(i), i\in {1,...,N_p}$ and $s_s(k), {k\in 1,...,N_s}$ are the symbols transmitted by the PU and SU transmitters and $n_{s(l)}$ is the noise at the SU receiver $l$. The SU performance can be quantified by the Signal-to-Noise-plus-Interference-Ratio (SINR) at receiver side. Considering (\ref{eq:1}), the SINR of the communication of the SU pair $l$, independent to the symbols realization, is calculated as
\begin{equation}\label{eq:2}
SINR_{s(l)}= \frac{\left\|\mathbf{w}_{l}^{H}\mathbf{H}_{ss(l,l)}\mathbf{m}_{l}\right\|^2}{\sum_{i=1}^{N_p}{\left\|\mathbf{w}_{l}^{H}\mathbf{h}_{ps(i,l)}\right\|^2} + \sum_{k=1;k\neq l}^{N_s}{\left\|\mathbf{w}_{l}^{H}\mathbf{H}_{ss(k,l)}\mathbf{m}_k\right\|^2} + \sigma_{s(l)}^2},
\end{equation}
\noindent where $\sigma_{s(l)}^2$ denotes the noise variance at the receiver in the SU pair $l$. The SINR at the SU receiver $l$ can be also rewritten in the following form
\begin{equation}\label{eq:3}
SINR_{s(l)}= \frac{\mathbf{m}_{l}^{H}\mathbf{H}^{H}_{ss(l,l)}\mathbf{w}_{l}\mathbf{w}^{H}_{l}\mathbf{H}_{ss(l,l)}\mathbf{m}_{l}}{\sum_{i=1}^{N_p}{\mathbf{h}_{ps(i,l)}^{H}\mathbf{w}_{l}\mathbf{w}_{l}^{H}\mathbf{h}_{ps(i,l)}} + \sum_{k=1;k\neq l}^{N_s}{\mathbf{m}_k^{H}\mathbf{H}_{ss(k,l)}^{H}\mathbf{w}_{l}\mathbf{w}_{l}^{H}\mathbf{H}_{ss(k,l)}\mathbf{m}_k} + \sigma_{s(l)}^2}.
\end{equation}
\noindent Let us define the following $N_t\text{x}N_t$ positive semidefinite matrices
\begin{equation}\label{eq:4}
\begin{gathered}
\mathbf{M}_k = \mathbf{m}_k\mathbf{m}_k^{H} \succ 0, \text{rank}(\mathbf{M}_k)=1, \forall k \in \left\{1,...,N_s\right\},\\
\mathbf{G}_{ss(k,l)} = \mathbf{H}_{ss(k,l)}^{H}\mathbf{w}_l\mathbf{w}_l^{H}\mathbf{H}_{ss(k,l)} \succ 0, \forall k,l \in \left\{1,...,N_s\right\},\\
\mathbf{G}_{sp(k,j)} = \mathbf{h}_{sp(k,j)}^{H}\mathbf{h}_{sp(k,j)} \succ 0, \text{rank}(\mathbf{G}_{sp(k,j)})=1, \forall k \in \left\{1,...,N_s\right\}, \forall j \in \left\{1,...,N_p\right\}.
\end{gathered}
\end{equation}
\noindent Substituting these matrix terms in (\ref{eq:3}), the SINR at the SU receiver $l$ can be calculated as
\begin{equation}\label{eq:5}
SINR_{s(l)}= \frac{\text{tr}(\mathbf{G}_{ss(l,l)}\mathbf{M}_l)}{\sum_{k=1;k\neq l}^{N_s}{\text{tr}(\mathbf{G}_{ss(k,l)}\mathbf{M}_k)} + I_{ps(l)} + \sigma_{s(l)}^2},
\end{equation}
\noindent where $I_{ps(l)}=\sum_{i=1}^{N_p}{\mathbf{h}_{ps(i,l)}^{H}\mathbf{w}_{l}\mathbf{w}_{l}^{H}\mathbf{h}_{ps(i,l)}}$ is the summary PU interference received at the SU receiver $l$ side. We further define the $N_tN_s\text{x}N_tN_s$ block diagonal matrices
\begin{equation}\label{eq:6}
\begin{gathered}
\mathbf{M} = \text{blkdiag}\left\{\mathbf{M}_1,...,\mathbf{M}_{N_s}\right\}, \text{rank}(\mathbf{M}) = N_s,\\
\mathbf{X}_{s(l)} = \text{blkdiag}\left\{\mathbf{G}_{ss(1,l)},...,\mathbf{G}_{ss(l-1,l)},\left[0\right]_{N_t\text{x}N_t},\mathbf{G}_{ss(l+1,l)},...,\mathbf{G}_{ss(N_s,l)}\right\}, l \in \left\{1,...,N_s\right\},\\
\mathbf{Q}_{s(l)} = \text{blkdiag}\left\{\left[0\right]_{N_t\text{x}N_t},...,\left[0\right]_{N_t\text{x}N_t},\mathbf{G}_{ss(l,l)},\left[0\right]_{N_t\text{x}N_t},...,\left[0\right]_{N_t\text{x}N_t}\right\}, l \in \left\{1,...,N_s\right\},\\
\mathbf{X}_{p(j)} = \text{blkdiag}\left\{\mathbf{G}_{sp(1,j)},...,\mathbf{G}_{sp(N_s,j)}\right\}, j \in \left\{1,...,N_p\right\},
\end{gathered}
\end{equation}
\noindent which conserve the positive semidefinite property of the comprising $N_t\text{x}N_t$ matrices. Here, $\left[0\right]_{N_t\text{x}N_t}$ denotes a zero matrix of size $N_t\text{x}N_t$.\\
\indent Based on the block diagonal matrices defined in (\ref{eq:6}), the SINR at the SU receiver $l$ can be simplified to be
\begin{equation}\label{eq:7}
SINR_{s(l)}= \frac{\text{tr}(\mathbf{Q}_{s(l)}\mathbf{M})}{\text{tr}(\mathbf{X}_{s(l)}\mathbf{M}) + I_{ps(l)} + \sigma_{s(l)}^2}.
\end{equation}
\noindent All the SU pairs now have a SINR dependence on the same matrix term $M$, which is a rank $N_s$ matrix, comprising all transmit beamforming vectors information. This derivation will prove to have a significant impact onto the convexity and resolvability of the optimal transmit beamforming vectors. 
\noindent Finally, the power constraints of the transmit and received beamformers can be defined as
\begin{equation}\label{eq:8}
\begin{gathered}
\left\|\mathbf{m}_l\right\|^2 = \text{tr}(\mathbf{M}_l) \leq P_t, \forall l \in \left\{1,...,N_s\right\},\\
\left\|\mathbf{w}_l\right\|^2 \leq 1, \forall l \in \left\{1,...,N_s\right\}.
\end{gathered}
\end{equation}
\noindent A typical characteristic in the multiuser MIMO IC channel is the power constraint per SU transmitter ($\leq P_t$), while the receiver beamformer is constrained to have a max. of unit norm.

\subsubsection {Primary system perspective}
\indent The signal received at the PU receiver $j$ comprises the signal coming from the PU transmitter $j$, the cumulative interference from the all other PU transmitters ($i\neq j$), the cumulative (a.k.a. aggregate) interference caused by all SU transmitters, multiplied by the respective channels, as well as the additive $\mathcal{CN}$ noise, i.e.
\begin{equation}\label{eq:9}
y_{p(j)}= h_{pp(j,j)}s_{p(j)} + \sum_{i=1; i\neq j}^{N_p}{h_{pp(i,j)}s_{p(j)}} + \sum_{k=1}^{N_s}{\mathbf{h}_{sp(k,j)}\mathbf{m}_ks_{s(k)}} + n_{p(j)},
\end{equation}
\noindent where $n_{p(j)}$ is the noise at the PU receiver $j$. Since the aim of the targeted algorithms in this paper is to operate in an underlay manner with the PU system, one of the metrics of interest is the cumulative SU interference power caused to the PU receiver $j$ (irrespective to the symbols' realization at the SU transmitter side). Utilizing the matrix terms from (\ref{eq:6}), we define the SU interference temperature constraint for all primary system receivers as
\begin{equation}\label{eq:10}
I_{sp(j)}= \sum_{k=1}^{N_s}{\left\|\mathbf{h}_{sp(k,j)}\mathbf{m}_k\right\|^2} = \text{tr}(\mathbf{X}_{p(j)}\mathbf{M}) \leq \gamma, \forall j \in \left\{1,...,N_p\right\}.
\end{equation}
\noindent This constraint limits the aggregate interference caused by all SU transmissions ($\forall l \in \left\{1,...,N_s\right\}$) to all primary receivers ($\forall j \in \left\{1,...,N_p\right\}$) below a predefined threshold $\gamma$. This is a generic underlay spectrum sharing primary system protection constraint.
\indent This subsection presented the respective derivations (SUs SINR and PU interference constraint) for the multiuser MIMO CR interference channel. Considering Fig.~\ref{fig:1a}, it is intuitive that there is no limitation on the number of operating secondary links $N_s$, i.e. supported concurrent data streams. The combined virtual MIMO system has $N_s\text{x}N_t$ transmit and $N_s\text{x}N_r$ receive antennas, meaning that the degrees of freedom in this system ranges with the number of SU pairs. 
\subsection {Multiuser MIMO CR broadcast and multiple-access channels}
\indent The multiuser MIMO broadcast and multiple-access channels are analogous to a scenario of single cell MIMO transmissions. A broadcast channel is the case of single base station (BS) serving multiple terminals in downlink (Fig.~\ref{fig:1a}), while a MAC channel is the case of multiple terminals communicating with a single BS in uplink (Fig.~\ref{fig:1b}). The following subsections present the required simplifications to adapt the approach to these system models. 

\subsubsection {BC channel}
\indent In the case of the broadcast channel (Fig.~\ref{fig:1b}), the SU system is composed of a signal source (transmitter) and multiple signal destinations (receivers). In this case, the SU signal and the SU interference to a single SU receiver arrive via the same channel, multiplied with the respective transmit beamforming vectors. The similar logic applies to the SU interference caused to the PU receivers. The following equations represent the differences (simplifications) with respect to the multiuser MIMO interference channel (eqs. (\ref{eq:4}) and (\ref{eq:6}))
\begin{equation}\label{eq:11}
\begin{gathered}
\mathbf{H}_{ss(1,l)} = \mathbf{H}_{ss(2,l)} = ... = \mathbf{H}_{ss(N_s,l)}, \forall l \in \left\{1,...,N_s\right\},\\
\mathbf{G}_{ss(1,l)} = \mathbf{G}_{ss(2,l)} = ... = \mathbf{G}_{ss(N_s,l)}, \forall l \in \left\{1,...,N_s\right\},\\
\mathbf{h}_{sp(1,j)} = \mathbf{h}_{sp(2,j)} = ... = \mathbf{h}_{sp(N_s,j)}, \forall j \in \left\{1,...,N_p\right\},\\
\mathbf{G}_{sp(1,j)} = \mathbf{G}_{sp(2,j)} = ... = \mathbf{G}_{sp(N_s,j)}, \forall j \in \left\{1,...,N_p\right\}.
\end{gathered}
\end{equation}
\noindent For the multiuser MIMO broadcast channel, the transmit power constraint in (\ref{eq:8}), is defined as
\begin{equation}\label{eq:12}
\sum_{l=1}^{N_s}{\left\|\mathbf{m}_l\right\|^2} = \text{tr}(\mathbf{M}) \leq P_t
\end{equation}
\noindent As can be noticed, the power is shared among the transmit beamformers (i.e. SU data streams), providing options for transmit power allocation optimization. Finally, the broadcast channel imposes a limitation of the number of served secondary data streams ($N_s$). Considering equation (\ref{eq:11}) and Fig.~\ref{fig:1b}, it is intuitive that the degrees of freedom are limited by the number of transmit antennas $N_t$. The combined virtual MIMO system has $N_t$ transmit and $N_s\text{x}N_r$ receive antennas, meaning that the number of served SU data streams must be $N_s \leq N_t$.

\subsubsection {MAC channel}
\indent In the case of the MAC channel (Fig.~\ref{fig:1c}), the SU system is composed of multiple signal sources (transmitters) and single signal destination (receiver). All SU signals and SU and PU interference arrive at the same receiver, imposing the following adaptations and simplifications with respect to the multiuser MIMO interference channel (eqs. (\ref{eq:4}) and (\ref{eq:6}))
\begin{equation}\label{eq:13}
\begin{gathered}
\mathbf{H}_{ss(k,1)} = \mathbf{H}_{ss(k,2)} = ... = \mathbf{H}_{ss(k,N_s)}, \forall k \in \left\{1,...,N_s\right\},\\
\mathbf{h}_{ps(i,1)} = \mathbf{h}_{ps(i,2)} = ... = \mathbf{h}_{ps(i,N_s)}, \forall i \in \left\{1,...,N_p\right\}.
\end{gathered}
\end{equation}
\noindent The transmit power constraint is the same as in case of the MIMO interference channel. Again, the MAC channel imposes a limitation of the number of served secondary data streams ($N_s$). Considering Equation (\ref{eq:13}) and Fig.~\ref{fig:1c}, it is clear that the degrees of freedom are limited by the number of receive antennas $N_r$. The combined virtual MIMO system has $N_s\text{x}N_t$ transmit and $N_r$ receive antennas, implying that the degrees of freedom in this system is limited by $N_r$. This means that the number of independent and concurrent secondary data streams must be $N_s \leq N_r$. 

\subsection{Summary}
\indent In general, the same SINR related equations (\ref{eq:2})-(\ref{eq:7}) and PU protection related expressions (\ref{eq:10}) can be used for all three models, with the respective simplifications shown in (\ref{eq:11}) and (\ref{eq:13}) for the multiuser BC and MAC models, respectively. Finally, a generic transmit power constraint for all considered channel models is defined as
\begin{equation}\label{eq:14}
\underbrace{\left\|\mathbf{m}_l\right\|^2 = \text{tr}(\mathbf{M}_l) \leq P_t, \forall l \in \left\{1,...,N_s\right\}}_{\textbf{IC and MAC channels}} \textbf{  or  } \underbrace{\sum_{l=1}^{N_s}{\left\|\mathbf{m}_l\right\|^2}=\text{tr}(\mathbf{M}) \leq P_t}_{\textbf{BC channel}},
\end{equation}
\noindent and the generic receive power constraint as
\begin{equation}\label{eq:15}
\left\|\mathbf{w}_l\right\|^2 \leq 1, \forall l \in \left\{1,...,N_s\right\}.
\end{equation}
\noindent Besides the previously denoted limitations on the number of served secondary data streams, all system models have a limitation on the number of protected PU receivers. In particular, in order to provide a secondary communication, the maximum number of protected primary receivers is limited by $N_t-1$, for each of the three multiuser MIMO system models, irrespective on the number of SU transmitters. This limitation is caused by the interference alignment capabilities, i.e. such system is able to provide transmit beamforming based  interference alignment of up to maximum of $N_t$ streams. Note that this limitation, along with the number of served SU data streams limitations are all system model limitations. All simplifications, limitations and scenario specifics are summarized in Table \ref{tab:1}. Further adaptations can be easily made to cover additional scenarios (system models), such as the multicell multiuser MIMO scenario. In the next section we will propose two generic multiuser beamforming strategies for underlay spectrum sharing.
\begin{table}[ht]
\small
\caption{Multiuser MIMO system model specifics for coordinated beamforming optimization}
\label{tab:1}
\centering
\begin{tabular}{|l||c||c||c|}
\hline
\textbf{System model} & \textbf{Simplifications} & \textbf{Limitations} & \textbf{Capabilities} \\
\hline
\textbf{Interference channel} & / & $N_p \leq N_t - 1$ & transmit/receive bf. \\
\hline
\multirow{2}{*}{}\textbf{Broadcast channel} & in (\ref{eq:11}) & $N_p \leq N_t - 1$ & transmit/receive bf. \\
& & $N_s \leq N_t$ &  power allocation \\
\hline
\multirow{2}{*}{}{\textbf{MAC channel}} & in (\ref{eq:13}) & $N_p \leq N_t - 1$ & transmit/receive bf. \\
& & $N_s \leq N_r$ & \\
\hline
\end{tabular}
\end{table}

\section {Generic multiuser beamforming algorithms for underlay sharing}
\label{sec:genbf}
\indent In this section we utilize our results from the previous section in order to design two generic multiuser beamforming algorithms for underlay spectrum sharing. In particular, the algorithms assume operation in a centralized manner, i.e. cooperation between the SU nodes. All channels are assumed to be frequency flat, fixed, and perfectly known (full channel state information) prior to the optimization. This is a justified assumption, since each of the channels can be estimated prior to the optimization and the operation. In this section we specifically will derive a \textit{fairness optimization algorithm} where the radio resources are shared among SUs to achieve balanced SINR distribution per secondary data stream and a \textit{sum rate maximization algorithm} that maximizes the sum rate of the secondary system. \\
\indent Since the problem of finding the optimal transmit and receive beamforming vectors simultaneously is a non-convex problem, both proposed algorithms treat the optimization iteratively. In particular, the general idea is to divide the transmit and the receive beamforming optimization into two separate sub-problems and optimize the beamforming vectors recursively, i.e. calculate the transmit beamformers for known (previously optimized) receive beamformers and vice versa. 

\subsection {Multiuser CR beamforming fairness algorithm}
\indent The multiuser beamforming fairness optimization for underlay spectrum sharing can be realized using SINR max-min optimization with respect to the primary system and power constraints. Utilizing the SINR expression in (\ref{eq:3}), the optimization problem can be formulated as
\begin{equation}\label{eq:16}
\begin{gathered}
\left\{\mathbf{m}_l,\mathbf{w}_l\right\}_{l=1}^{N_s} = \arg\max\limits_{\left\{\mathbf{m}_l,\mathbf{w}_l\right\}_{l=1}^{N_s}}\min_{l}\left\{SINR_{s(l)}\right\},\\
\text{subject to: (\ref{eq:10}), (\ref{eq:14}) and (\ref{eq:15})}.
\end{gathered}
\end{equation}
\noindent In particular, this optimization problem tends to balance the SINR distribution per SU data stream, keeping the aggregate interference caused to all the primary receivers below a predefined threshold $\gamma$, and the transmit power of SU transmitter(s) below a maximum of $P_t$. The receive beamformers have a unit norm constraint. However, the simultaneous optimization of both the receive and transmit beamformers, results in a non-convex optimization problem. In order to transform the problem into a convex one, we divide the optimization problem into two sub-problems, i.e. the transmit beamforming and the receive optimization sub-problems.

\subsubsection {Transmit beamforming optimization sub-problem}
\indent Assuming that the receive beamforming vectors are known, a transmit beamforming fairness optimization sub-problem can be formulated using the SINR definition in (\ref{eq:7}) and the respective constraints in (\ref{eq:10}) and (\ref{eq:14}). This yields
\begin{equation}\label{eq:17}
\begin{gathered}
\mathbf{M} = \arg\max\limits_{\mathbf{M}}\min_{l}\left\{\frac{\text{tr}(\mathbf{Q}_{s(l)}\mathbf{M})}{\text{tr}(\mathbf{X}_{s(l)}\mathbf{M}) + I_{ps(l)} + \sigma_{s(l)}^2}\right\},\\
\text{subject to: }\text{rank}(\mathbf{M}_l) = 1, \forall l \in \left\{1,...,N_s\right\},\text{ (\ref{eq:10}) and (\ref{eq:14})}.
\end{gathered}
\end{equation}
\noindent The problem in (\ref{eq:17}) can be transformed into a convex one with a semidefinite relaxation, i.e. omitting the rank constraints. It can be solved with \textit{Dinkelbach} fractional programming type of algorithms \cite{cite41} and using common semidefinite optimization tools. The aim is to find the positive semidefinite block diagonal non-full rank $N_tN_s\text{x}N_tN_s$ matrix $\mathbf{M}$, and extract the optimal transmit beamforming vectors accordingly in each iteration of the algorithm. This iteration based Dinkelbach algorithm in the $n$-th iteration operates on the following optimization problem (which is equivalent to a semidefinite relaxed problem in (\ref{eq:17})):
\begin{equation}\label{eq:18}
\begin{gathered}
\mathbf{M^{(\mathit{n})}} = \arg\max\limits_{\mathbf{M}}(\tau)\\
\text{subject to: } \text{tr}(\mathbf{Q}_{s(l)}\mathbf{M}) - \delta_{min}(\text{tr}(\mathbf{X}_{s(l)}\mathbf{M}) + I_{ps(l)} + \sigma_{s(l)}^2) \geq \tau,\text{ (\ref{eq:10}) and (\ref{eq:14})}, \\
\end{gathered}
\end{equation}
where $\delta_{min} = \min\left\{SINR_{s(1)}^{(n-1)},...,SINR_{s(N_s)}^{(n-1)}\right\}$ is the minimal SINR acquired in the previous $n-1$ iteration of the algorithm, and $\tau$ represents the fairness coefficient, i.e. the difference between the maximal and the minimal $SINR$. The transmit beamformers $\left\{\mathbf{m}_l\right\}_{l=1}^{N_s}$, are calculated as the principal singular vectors of the positive semidefinite matrices $\left\{\mathbf{M}_l\right\}_{l=1}^{N_s}$ extracted from the block diagonal matrix $\mathbf{M}$ obtained by the transmit beamforming optimization in (\ref{eq:18}).

\subsubsection {Receive beamforming optimization sub-problem}
\indent The optimal unit norm receive beamforming vectors $\left\{\mathbf{w}_l\right\}_{l=1}^{N_s}$, that maximize the SINR per SU data stream $l$ (eq. (\ref{eq:7})), for known transmit beamforming vectors $\left\{\mathbf{m}_l\right\}_{l=1}^{N_s}$, can be calculated as a Rayleigh quotient:
\begin{equation}\label{eq:19}
\mathbf{w}_l = \frac{\left(\sum_{i=1}^{N_p}{\mathbf{h}_{ps(i,l)}\mathbf{h}_{ps(i,l)}^{H}} + \sum_{k=1;k\neq l}^{N_s}{\mathbf{H}_{ss(k,l)}\mathbf{m}_k\mathbf{m}_k^{H}\mathbf{H}_{ss(k,l)}^{H}} + \sigma_{s(l)}^2\mathbf{I}_{N_r}\right)^{-1}\mathbf{H}_{ss(l,l)}\mathbf{m}_l}{\left\|\left(\sum_{i=1}^{N_p}{\mathbf{h}_{ps(i,l)}\mathbf{h}_{ps(i,l)}^{H}} + \sum_{k=1;k\neq l}^{N_s}{\mathbf{H}_{ss(k,l)}\mathbf{m}_k\mathbf{m}_k^{H}\mathbf{H}_{ss(k,l)}^{H}} + \sigma_{s(l)}^2\mathbf{I}_{N_r}\right)^{-1}\mathbf{H}_{ss(l,l)}\mathbf{m}_l\right\|}
\end{equation}

\subsubsection{Algorithm for multiuser CR beamforming fairness optimization}
\indent The transmit and receive optimizations presented in previous subsections, can be programmed as a recursive fairness optimization algorithm in the following manner. The optimal beamforming vectors are calculated in a cyclic manner, i.e. calculating the transmit beamformers for known receive beamformers, and vice versa, calculating receive beamformers for known transmit beamformers until iteratively reaching a globally optimal solution. Algorithm \ref{alg:1} presents the pseudo code of this multiuser CR beamforming fairness optimization algorithm. 
\begin{algorithm}
\small
\caption{\small Multiuser CR beamforming fairness optimization}\label{alg:1}
\textbf{STEP 1: Initialization}
\begin{algorithmic}
\STATE{$n\gets 1$}
\FOR{$l=1:N_s$}
\STATE{$\mathbf{m}_{l}^{(n)},\mathbf{w}_{l}^{(n)}\gets$ random feasible vectors}
\STATE{$SINR_{s(l)}^{(n)}\gets$ based on (\ref{eq:3})}
\ENDFOR
\end{algorithmic}
\textbf{STEP 2: Minimum SINR calculation}
\begin{algorithmic}
\STATE{$n\gets n+1$}
\STATE{$\delta_{min}\gets \min{\{SINR_{s(1)}^{(n-1)},...,SINR_{s(N_s)}^{(n-1)}\}}$}
\end{algorithmic}
\textbf{STEP 3: Transmit beamforming optimization}
\begin{algorithmic}
\STATE{$\left[\mathbf{M}^{(n)},\tau\right]\gets$ based on (\ref{eq:18})}
\FOR{$l=1:N_s$}
\STATE{$\mathbf{m}_{l}^{(n)} \gets$ principal singular vector of $\mathbf{M}_{l}^{(n)}$}
\ENDFOR
\end{algorithmic}
\textbf{STEP 4: Receive beamforming optimization}
\begin{algorithmic}
\FOR{$l=1:N_s$}
\STATE{$\mathbf{w}_{l}^{(n)}\gets$ based on (\ref{eq:19})}
\ENDFOR
\end{algorithmic}
\textbf{STEP 5: Stopping criteria}
\begin{algorithmic}
\IF{$\tau > \epsilon$}
\FOR{$l=1:N_s$}
\STATE{$SINR_{s(l)}^{(n)}\gets$ based on (\ref{eq:3})}
\ENDFOR
\STATE{go back to \textbf{STEP 2}} 
\ELSE
\FOR{$l=1:N_s$}
\STATE{$\mathbf{m}_{l} \gets \mathbf{m}_{l}^{(n)}$; $\mathbf{w}_{l} \gets \mathbf{w}_{l}^{(n)}$}
\ENDFOR
\STATE{\textbf{Stop algorithm}}
\ENDIF
\end{algorithmic}
\end{algorithm}

\subsection {Multiuser CR beamforming sum rate maximization algorithm}
\indent Employing Shannon capacity, the achievable capacities for the SU data stream $l$ and the entire secondary system are
\begin{equation}\label{eq:20}
\begin{gathered}
C_{s(l)}= \log_2(1+SINR_{s(l)}),\\
C_s = \sum_{l=1}^{N_s}{C_{s(l)}}.
\end{gathered}
\end{equation}
\indent Based on the SINR expression in (\ref{eq:3}) the beamforming optimization problem for the underlay sharing can be defined to maximize the sum rate of the secondary data streams complying with the interference constraints, i.e.
\begin{equation}\label{eq:21}
\begin{gathered}
\left\{\mathbf{m}_l,\mathbf{w}_l\right\}_{l=1}^{N_s} = \arg\max\limits_{\left\{\mathbf{m}_l,\mathbf{w}_l\right\}_{l=1}^{N_s}}\left\{\sum_{l=1}^{N_s}{\log_2(1+SINR_{s(l)})}\right\},\\
\text{subject to: (\ref{eq:10}), (\ref{eq:14}) and (\ref{eq:15})}.
\end{gathered}
\end{equation}
\noindent In this optimization we aim to maximize the summary secondary system sum rate, while keeping the aggregate interference caused to all the primary receivers below a predefined threshold $\gamma$, the transmit power of all SU transmitter(s) below a maximum of $P_t$ and the receive beamformers power below one. The Equation (\ref{eq:21}) is a non-convex optimization problem. Therefore, we consider several techniques to alleviate the non-convexity of the optimization problem and transform the problem into a convex one. Similarly to the fairness optimization, we divide the problem into two separate sub-problems and optimize the beamforming vectors recursively.

\subsubsection{Transmit beamforming optimization sub-problem}
\indent In order to transform the problem in (\ref{eq:21}) into a convex one, as mentioned, first, the problem is considered only from the transmit beamforming optimization perspective, utilizing the SINR expression in (\ref{eq:7}) and performing a semidefinite relaxation
\begin{equation}\label{eq:22}
\begin{gathered}
\mathbf{M} = \arg\max\limits_{\mathbf{M}}\left\{\sum_{l=1}^{N_s}{\log_2\left(1+\frac{\text{tr}(\mathbf{Q}_{s(l)}\mathbf{M})}{\text{tr}(\mathbf{X}_{s(l)}\mathbf{M}) + I_{ps(l)} + \sigma_{s(l)}^2}\right)}\right\}\\
\text{subject to:  (\ref{eq:10}) and (\ref{eq:14})}.
\end{gathered}
\end{equation}
\noindent The sum rate of the secondary data streams can be represented as difference between two sums of logarithmic terms
\begin{equation}\label{eq:23}
\begin{gathered}
C_s = f(\mathbf{M})-g(\mathbf{M}),\\
f(\mathbf{M}) = \sum_{l=1}^{N_s}{\log_2\left(\text{tr}\left((\mathbf{Q}_{s(l)}+\mathbf{X}_{s(l)})\mathbf{M}\right) + I_{ps(l)} + \sigma_{s(l)}^2\right)},\\
g(\mathbf{M}) = \sum_{l=1}^{N_s}{\log_2\left(\text{tr}\left(\mathbf{X}_{s(l)}\mathbf{M}\right) + I_{ps(l)} + \sigma_{s(l)}^2\right).}
\end{gathered}
\end{equation}
\noindent Both terms, $f(\mathbf{M})$ and $g(\mathbf{M})$, are concave functions, and therefore, the optimization problem in (\ref{eq:22}) is still a non-convex one. In order to alleviate the non-convexity we use the same approach as in \cite{cite39}. In particular, due to the properties of the logarithm function, $g(\mathbf{M})$ is weakly sensitive to changes of the variable $M$, so in a local (and fairly large) neighborhood of a random point $\mathbf{M^{(\mathit{n})}}$ we can use its linear (first order Taylor) approximation
\begin{equation}\label{eq:24}
g(\mathbf{M}) \approx g(\mathbf{M^{(\mathit{n})}}) + \left\langle \nabla g(\mathbf{M^{(\mathit{n})}}), \mathbf{M}-\mathbf{M^{(\mathit{n})}}\right\rangle,
\end{equation}
\noindent where $\nabla g(\mathbf{M})$ is the first order derivative of the function $g(\mathbf{M})$ in point $\mathbf{M^{(\mathit{n})}}$, calculated as
 \begin{equation}\label{eq:25}
\nabla g(\mathbf{M}) = \sum_{l=1}^{N_s}{\frac{\mathbf{X}_{s(l)}}{\left(\text{tr}\left(\mathbf{X}_{s(l)}\mathbf{M}\right) + I_{ps(l)} + \sigma_{s(l)}^2\right)\ln{2}}}.
\end{equation}
\noindent Considering (\ref{eq:23}), (\ref{eq:24}) and (\ref{eq:25}) the sum rate function in the local neighborhood of $\mathbf{M^{(\mathit{n})}}$ can be approximated as
\begin{equation}\label{eq:26}
C_s \approx f(\mathbf{M}) - g(\mathbf{M^{(\mathit{n})}}) - \left\langle \nabla g(M^{(\mathit{n})}), \mathbf{M}-\mathbf{M^{(\mathit{n})}}\right\rangle,
\end{equation}
\noindent which is now a concave function and has a single optimum $\mathbf{M^{(\mathit{n}+\text{1})}}$. In order to derive a sum rate optimization problem that will globally converge, we have to consider an important property of the function $g(\mathbf{M})$. Since this function is concave, its gradient is also its super gradient, i.e. $g(\mathbf{M})$ has the following property
\begin{equation}\label{eq:27}
g(\mathbf{M}) \leq g(\mathbf{M^{(\mathit{n})}}) + \left\langle \nabla g(\mathbf{M^{(\mathit{n})}}), \mathbf{M}-\mathbf{M^{(\mathit{n})}}\right\rangle.
\end{equation}
\noindent This property results in the following corollary
\begin{equation}\label{eq:28}
\begin{split}
f(\mathbf{M^{(\mathit{n}+\text{1})}}) - g(\mathbf{M^{(\mathit{n}+\text{1})}}) & \geq f(\mathbf{M^{(\mathit{n})}}) - g(\mathbf{M^{(\mathit{n})}}) - \left\langle \nabla g(\mathbf{M^{(\mathit{n})}}), \mathbf{M}-\mathbf{M^{(\mathit{n})}}\right\rangle\\
 & \geq f(\mathbf{M^{(\mathit{n})}}) - g(\mathbf{M^{(\mathit{n})}}).
\end{split}
\end{equation}
\noindent This means that the new solution $\mathbf{M^{(\mathit{n}+\text{1})}}$ is always better than the previous $\mathbf{M^{(\mathit{n})}}$. Since that constraint set in (\ref{eq:22}) is compact, starting from a random feasible point, the optimization will always find a better feasible point. The sum rate optimization problem in (\ref{eq:22}) can be now solved iteratively, with the optimal $\mathbf{M^{(\mathit{n})}}$ in the $n$-th iteration calculated as an output of the following convex optimization: 
\begin{equation}\label{eq:29}
\begin{gathered}
\mathbf{M^{(\mathit{n})}} = \arg\max\limits_{\mathbf{M}}\left\{f(\mathbf{M}) - g(\mathbf{M^{(\mathit{n}-\text{1})}}) - \left\langle \nabla g(\mathbf{M^{(\mathit{n}-\text{1})}}), \mathbf{M}-\mathbf{M^{((\mathit{n}-\text{1})}}\right\rangle\right\},\\
\text{subject to: (\ref{eq:10}) and (\ref{eq:14})}.
\end{gathered}
\end{equation}
\indent Similar to the fairness algorithm, the transmit beamformers $\left\{\mathbf{m}_l\right\}_{l=1}^{N_s}$ are calculated as the principal singular vectors of the positive semidefinite matrices $\left\{\mathbf{M}_l\right\}_{l=1}^{N_s}$ extracted from the block diagonal matrix $\mathbf{M}$ obtained by the transmit beamforming optimization in (\ref{eq:22}).

\subsubsection{Receive beamforming optimization sub-problem}
\indent The unit norm receive beamforming vectors $\left\{\mathbf{w}_l\right\}_{l=1}^{N_s}$, that maximize the SINR per SU data stream is calculated in the same manner as in the fairness algorithm, using formula (\ref{eq:19}).

\subsubsection{Algorithm for multiuser CR beamforming fairness optimization}
\indent The recursive (transmit/receive beamforming) sum rate optimization algorithm for underlay spectrum sharing is defined in the following manner. The optimal beamforming vectors are calculated in a iterative manner, i.e. calculating the transmit beamformers for known receive beamformers, and vice versa, calculating receive beamformers for known transmit beamformers, until reaching a globally optimal solution. Algorithm \ref{alg:2} presents the pseudo code of this multiuser CR beamforming sum rate optimization algorithm. \\
\begin{algorithm}
\small
\caption{\small Multiuser CR beamforming sum rate optimization}\label{alg:2}
\textbf{STEP 1: Initialization}
\begin{algorithmic}
\STATE{$n\gets 1$}
\FOR{$l=1:N_s$}
\STATE{$\mathbf{m}_{l}^{(n)},\mathbf{w}_{l}^{(n)}\gets$ random feasible vectors}
\STATE{$SINR_{s(l)}^{(n)}\gets$ based on (\ref{eq:3})}
\ENDFOR
\STATE{$\mathbf{M}^{(n)} \gets$ based on (\ref{eq:4}) and (\ref{eq:6})} 
\STATE{$C_{s}^{(n)}\gets$ based on (\ref{eq:20})}
\end{algorithmic}
\textbf{STEP 2: $g, \nabla g$ calculation}
\begin{algorithmic}
\STATE{$n\gets n+1$}
\STATE{$g(\mathbf{M^{(\mathit{n}-\text{1})}})\gets$ based on (\ref{eq:23})}
\STATE{$\nabla g(\mathbf{M^{(\mathit{n}-\text{1})}})\gets$ based on (\ref{eq:25})}
\end{algorithmic}
\textbf{STEP 3: Transmit beamforming optimization}
\begin{algorithmic}
\STATE{$\mathbf{M}^{(n)}\gets$ based on (\ref{eq:29})}
\FOR{$l=1:N_s$}
\STATE{$\mathbf{m}_{l}^{(n)} \gets$ principal singular vector of $\mathbf{M}_{l}^{(n)}$}
\ENDFOR
\end{algorithmic}
\textbf{STEP 4: Receive beamforming optimization}
\begin{algorithmic}
\FOR{$l=1:N_s$}
\STATE{$\mathbf{w}_{l}^{(n)}\gets$ based on (\ref{eq:19})}
\ENDFOR
\end{algorithmic}
\textbf{STEP 5: Stopping criteria}
\begin{algorithmic}
\FOR{$l=1:N_s$}
\STATE{$SINR_{s(l)}^{(n)}\gets$ based on (\ref{eq:3})}
\ENDFOR
\STATE{$C_{s}^{(n)}\gets$ based on (\ref{eq:20})}
\IF{$(C_{s}^{(n)}-C_{s}^{(n-1)}) > \epsilon$} 
\STATE{go back to \textbf{STEP 2}} 
\ELSE
\FOR{$l=1:N_s$}
\STATE{$\mathbf{m}_{l} \gets \mathbf{m}_{l}^{(n)}$; $\mathbf{w}_{l} \gets \mathbf{w}_{l}^{(n)}$}
\ENDFOR
\STATE{\textbf{Stop algorithm}}
\ENDIF
\end{algorithmic}
\end{algorithm}
\indent In order to reduce the complexity of the recursive transmit beamforming optimization and the overall complexity of the multiuser CR beamforming underlay sharing (Algorithms \ref{alg:1} and \ref{alg:2}), this paper considers an additional simplification of the transmit beamforming sub-problem. Applying the generalized Kantorovich and Cauchy Schwartz inequalities, the SINR in eq.(\ref{eq:2}), can be easily shown to be lower bounded only by the transmit beamformers, i.e.
\begin{equation}\label{eq:30}
\begin{gathered}
SINR_{s(l)}= \frac{\mathbf{m}_{l}^{H}\mathbf{H}^{H}_{ss(l,l)}\mathbf{w}_{l}\mathbf{w}^{H}_{l}\mathbf{H}_{ss(l,l)}\mathbf{m}_{l}}{\sum_{i=1}^{N_p}{\mathbf{h}_{ps(i,l)}^{H}\mathbf{w}_{l}\mathbf{w}_{l}^{H}\mathbf{h}_{ps(i,l)}} + \sum_{k=1;k\neq l}^{N_s}{\mathbf{m}_k^{H}\mathbf{H}_{ss(k,l)}^{H}\mathbf{w}_{l}\mathbf{w}_{l}^{H}\mathbf{H}_{ss(k,l)}\mathbf{m}_k} + \sigma_{s(l)}^2}\\
\geq \frac{\mathbf{m}_{l}^{H}\mathbf{H}^{H}_{ss(l,l)}\mathbf{H}_{ss(l,l)}\mathbf{m}_{l}}{\sum_{i=1}^{N_p}{\mathbf{h}_{ps(i,l)}^{H}\mathbf{h}_{ps(i,l)}} + \sum_{k=1;k\neq l}^{N_s}{\mathbf{m}_k^{H}\mathbf{H}_{ss(k,l)}^{H}\mathbf{H}_{ss(k,l)}\mathbf{m}_k} + \sigma_{s(l)}^2} = SINR_{s(l)}^{tbf}.
\end{gathered}
\end{equation}
\noindent Considering that $SINR_{s(l)}$ is lower bounded by $SINR_{s(l)}^{tbf}$, we can do the transmit beamforming optimization just based on the based on the $SINR_{s(l)}^{tbf}$ expressions. Therefore, we define the following simplified terms
\begin{equation}\label{eq:31}
\begin{gathered}
I_{ps(l)}^{tbf} =\sum_{i=1}^{N_p}{\mathbf{h}_{ps(i,l)}^{H}\mathbf{h}_{ps(i,l)}},\\
\mathbf{G}_{ss(k,l)}^{tbf} = \mathbf{H}_{ss(k,l)}\mathbf{H}_{ss(k,l)}^{H} \succ 0, \forall k,l \in \left\{1,...,N_s\right\},\\
\mathbf{X}_{s(l)}^{tbf} = \text{blkdiag}\left\{\mathbf{G}_{ss(1,l)}^{tbf},...,\mathbf{G}_{ss(l-1,l)}^{tbf},\left[0\right]_{N_t\text{x}N_t},\mathbf{G}_{ss(l+1,l)}^{tbf},...,\mathbf{G}_{ss(N_s,l)}^{tbf}\right\} \succ 0, \forall l \in \left\{1,...,N_s\right\},\\
\mathbf{Q}_{s(l)}^{tbf} = \text{blkdiag}\left\{\left[0\right]_{N_t\text{x}N_t},...,\left[0\right]_{N_t\text{x}N_t},\mathbf{G}_{ss(l,l)}^{tbf},\left[0\right]_{N_t\text{x}N_t},...,\left[0\right]_{N_t\text{x}N_t}\right\} \succ 0, \forall l \in \left\{1,...,N_s\right\}.
\end{gathered}
\end{equation}
\noindent Now the transmit beamforming optimization can be performed irrespective to the receive beamforming, using these terms in the optimization problems provided by (\ref{eq:18}) and (\ref{eq:28}). The receive beamforming optimization is solved successively after the transmit beamforming optimization is finished. This will reduce the the number of matrix operations per iteration (refer to Algorithms \ref{alg:1} and \ref{alg:2}), but will result in sub-optimal behavior.

\subsection{Summary}
\indent This section presented two multiuser beamforming algorithms for underlay operation in primary system frequency bands, i.e. fairness and sum rate maximization algorithms, referred to as Fairness and SRM in the remaining text, respectively. Both algorithms can be applied to all three MIMO system models, irrespective on the number of antennas in the system. Although the BC and MAC system models impose limitation on the number of served secondary data streams $N_s$, the sum rate maximization algorithm can also work with higher number of secondary users, optimally selecting the subset of data streams with lowest channel correlation. Thus, this algorithm opportunistically maximizes the overall sum rate of the secondary system. In these cases, the fairness optimization algorithm will significantly decrease the overall sum rate (the rate per user). The following section will present the performance evaluation and prove the benefits of the employment of these algorithms in representative scenarios.

\section{Performance analysis}
\label{sec:PerfAnl}
In this section we assess the performance of the proposed beamforming algorithms by focussing on fairness and sum rate maximization. The performance analysis is performed in terms of the \textit{sum rate} (${C_s} = \sum\limits_{l = 1}^{Ns} {{{\log }_2}(1 + SIN{R_{s(l)}})}$) and the \textit{normalized sum rate} (${\Theta _s} = {C_s}/{C_{bound}}$)\footnote[1]{$C_{bound}$ denotes the sum rate \textit{point-to-point outer bound} of the $N_s$ active data streams, which assumes \textit{no SU and no PU interference} in the communication and can be computed as defined in ~\cite{cite35}. In order to provide easier tractability, for the remainder of the paper the term $C_{bound}$ will be denoted as the upper bound.}. We investigate also the benefits of exploiting the complexity reduction approach for successive optimization for both beamforming algorithms in terms of the \textit{Convergence rate}. Moreover, the section introduces the \textit{SNR deviation} (i.e. signal deviation, log-normally modeled) as an evaluation metric. The SNR deviation is a valuable metric that reflects the realistic behavior of the system. In practice it is almost impossible for all or group of secondary data streams to experience equal SNRs on their communication channels. The assumption of constant SNR at the users has been frequently utilized in previous works and can lead to significantly incorrect conclusions. Our scenario parameters are given in Table~\ref{tab:1a}.
\begin{table}[ht]
\small
\caption{Scenario Parameters (Performance analysis)}
\label{tab:1a}
\centering
\begin{tabular}{|c||c|}
\hline
\textbf{Parameters} & \textbf{Values} \\
\hline
Simulation environment & Matlab\\ 
\hline
Optimization tool & SDP via CVX \\
\hline
Stopping threshold ($\epsilon$) & $10^{-2}$ \\
\hline
$SNR$ [$dB$] & $0:30$ \\
\hline
SNR deviation ($\sigma$) [$dB$] & $0:5:10$ \\
\hline
Channel model & Circular complex normal \\
\hline
Transmit antennas $N_t$ & $2:2:6$ \\
\hline
Receive antennas $N_r$ & $2:2:6$ \\
\hline
PUI threshold ($\gamma$) [$mW$] & $10^{-10}$ \\
\hline
Number of PUs & $1:1:\left({N_t-1}\right)$ \\
\hline
Monte Carlo trials per configuration & $1000$ \\
\hline
\end{tabular}
\end{table}

\subsection{Interference channel}
\begin{figure*}[!t]
\centering
\subfloat[]{\includegraphics[width=3.4in, height=2.1in]{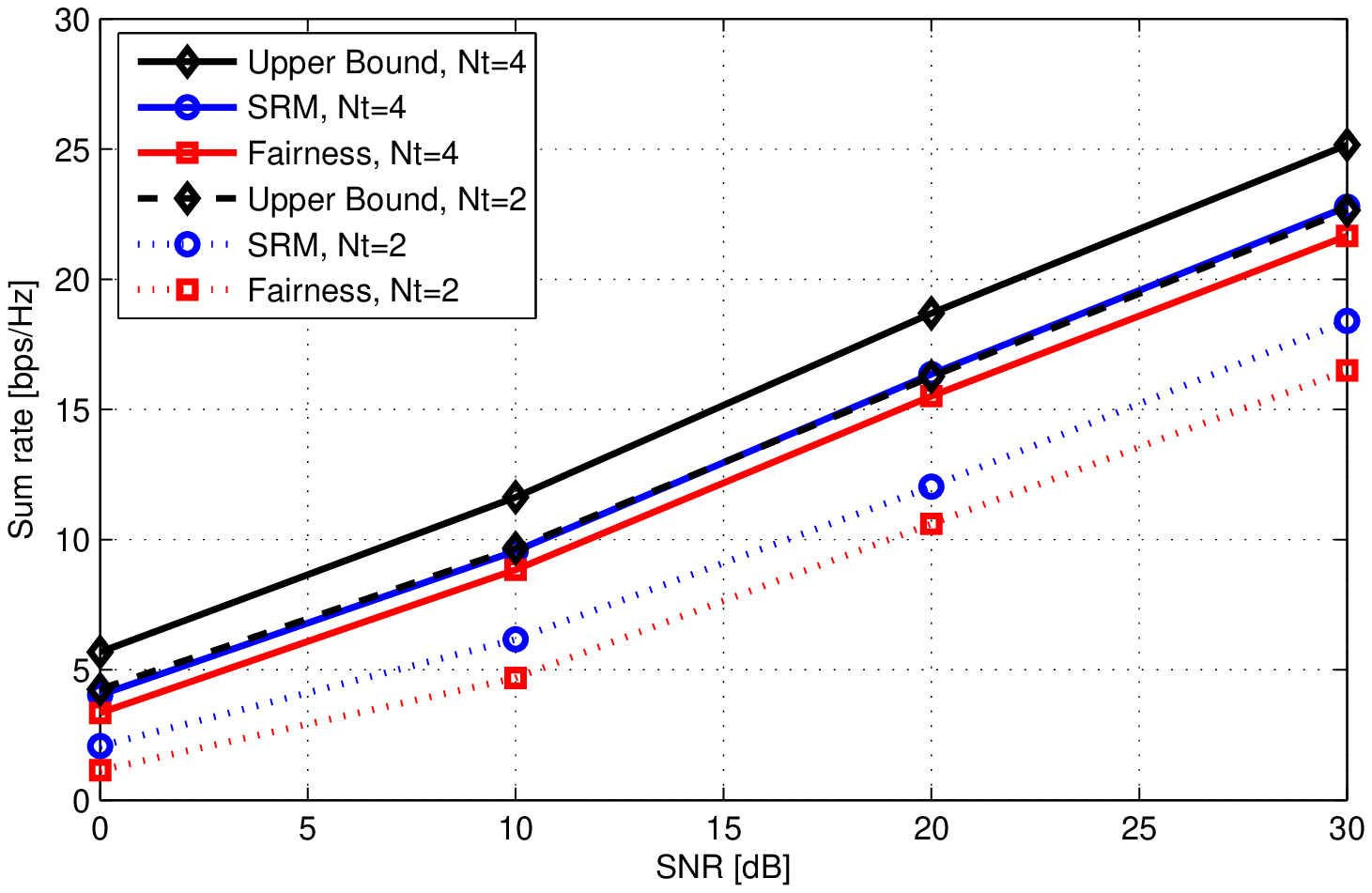}\label{fig:4.1}}
\subfloat[]{\includegraphics[width=3.4in, height=2.1in]{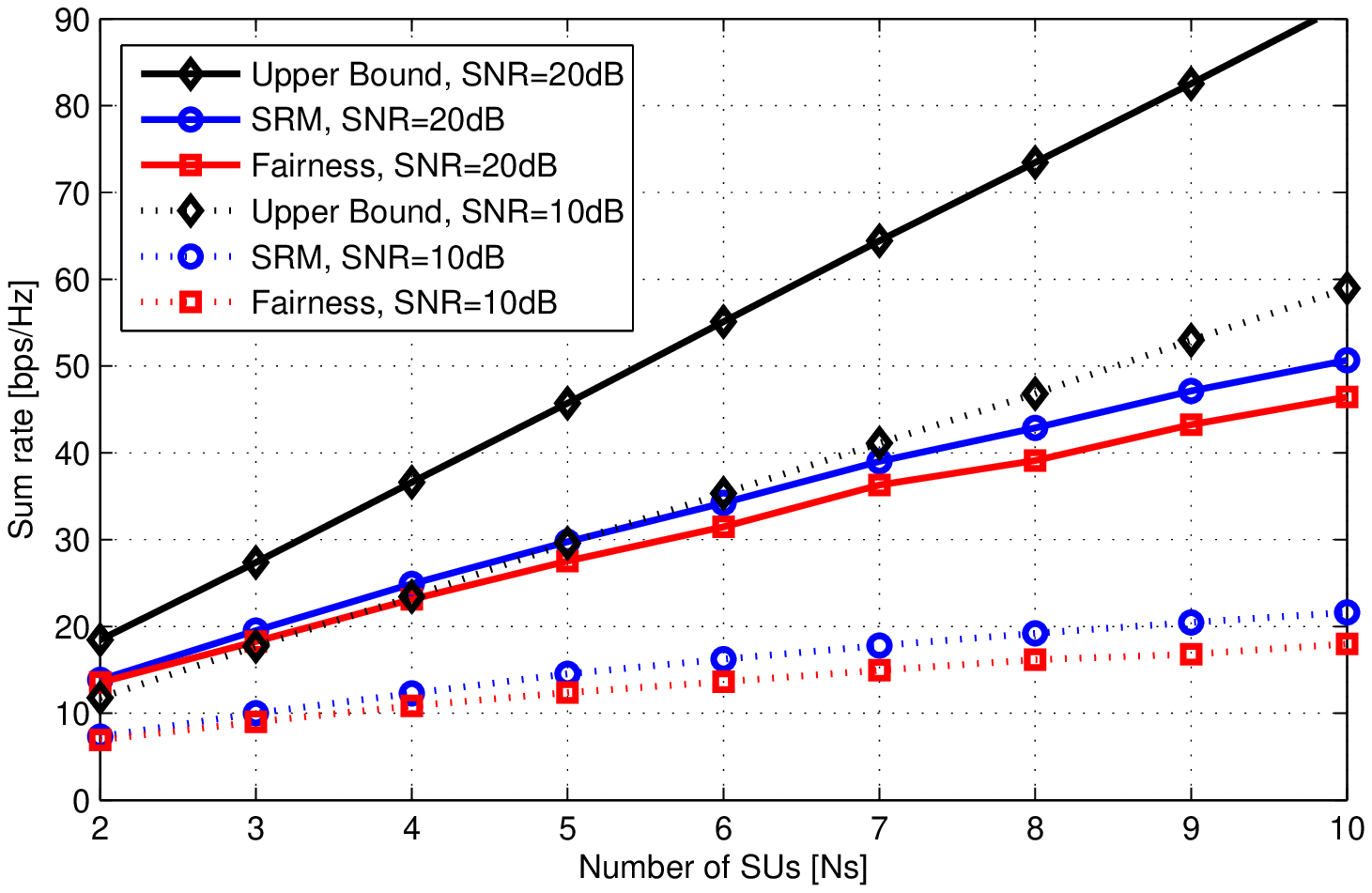}\label{fig:4.2}}
\caption{a) Sum rate vs SNR, for different antenna configurations ($N_r=2$, $N_s=2$, $N_p=1$, $\sigma=5dB$) b) Sum rate vs Number of SU devices, for different SNR values ($N_t=4$ $N_r=2$, $N_p=2$, $\sigma=0dB$)}
\end{figure*}

From system model and design perspective, the interference channel represents the most generic and complex scenario where all active pairs interfere between each other. In order to provide efficient communication over this type of channel, there must exist cooperation between the active nodes in the network. From the practical perspective, this scenario can be related to coordinated multi-cell communication as well as ad-hoc networks where multiple independent and physically collocated transmissions occur at the same time frame. \\
\indent Fig.~\ref{fig:4.1} depicts the achieved sum rate of the secondary system for the fairness and sum rate maximization (SRM) beamforming algorithms versus the SNR. We consider also different antenna configurations. It is evident that the SRM algorithm always outperforms the Fairness algorithm for any SNR and antenna configuration. However, the performance gain of the SRM over the Fairness algorithm diminishes for higher number of antennas. This behavior is a result of the higher number of available degrees of freedom in the system, facilitating improved sum rate performance per user and thus more efficient operation of the Fairness algorithm. In addition, Fig.~\ref{fig:4.1} compares the performance of both algorithms with respect to the upper bound. By increasing the number of antennas in the system, i.e. the degrees of freedom, both algorithms improve their performances and operate with rates that approach the upper bound.

\indent Fig.~\ref{fig:4.2} depicts the sum rate of the secondary system versus the number of active secondary users and different SNR values. The SRM algorithm always achieves higher sum rates compared to the Fairness algorithm. Moreover, for higher number of secondary users the performance gap between the beamforming algorithms increases. This is a result of the specific optimization goal of the underlying beamforming process. For the SRM case, the algorithm favors and primarily maximizes the sum rate of the data streams with favorable channels conditions (those that experience a lower PU and cross SU interference, higher SNR, etc.) neglecting the remaining data streams' rates. In contrast, the Fairness algorithm strives to leverage the performance of all users and induce fairness among them, regardless of the attained channels conditions. For high number of secondary users the amount of pairs with unfavorable channels conditions will increase, thus forcing the Fairness algorithm to improve the sum rates of those pairs at the price of decreasing the overall sum rate.  

\indent It is also evident from Fig.~\ref{fig:4.2} that both algorithms can serve arbitrarily large number of users, and are not bounded by the number of transmit and receive antennas. By carefully analyzing (\ref{eq:7}) and Fig.~\ref{fig:1a}, it is apparent that the solution space dimension scales with $N_s$ thus, providing the possibility to serve any number of secondary users. Nevertheless, the sum rate of both algorithms decreases in comparison to the upper bound as $N_s$ increases. This conclusion is counter intuitive with respect to the model presented in Equation (\ref{eq:7}), where the optimization process is centralized and exploits the information of all SU channels in the system. However, in terms of the transmission strategy for the secondary user streams, the optimization process is restricted only to take into consideration and utilize the antennas of the respective transmitter receiver pair. This limits the possible solution set for the optimization process and results in lower performance when compared with the upper bound.     

\subsection{Broadcast channel}
\begin{figure*}[!t]
\centering
\subfloat[]{\includegraphics[width=2.2in, height=1.7in]{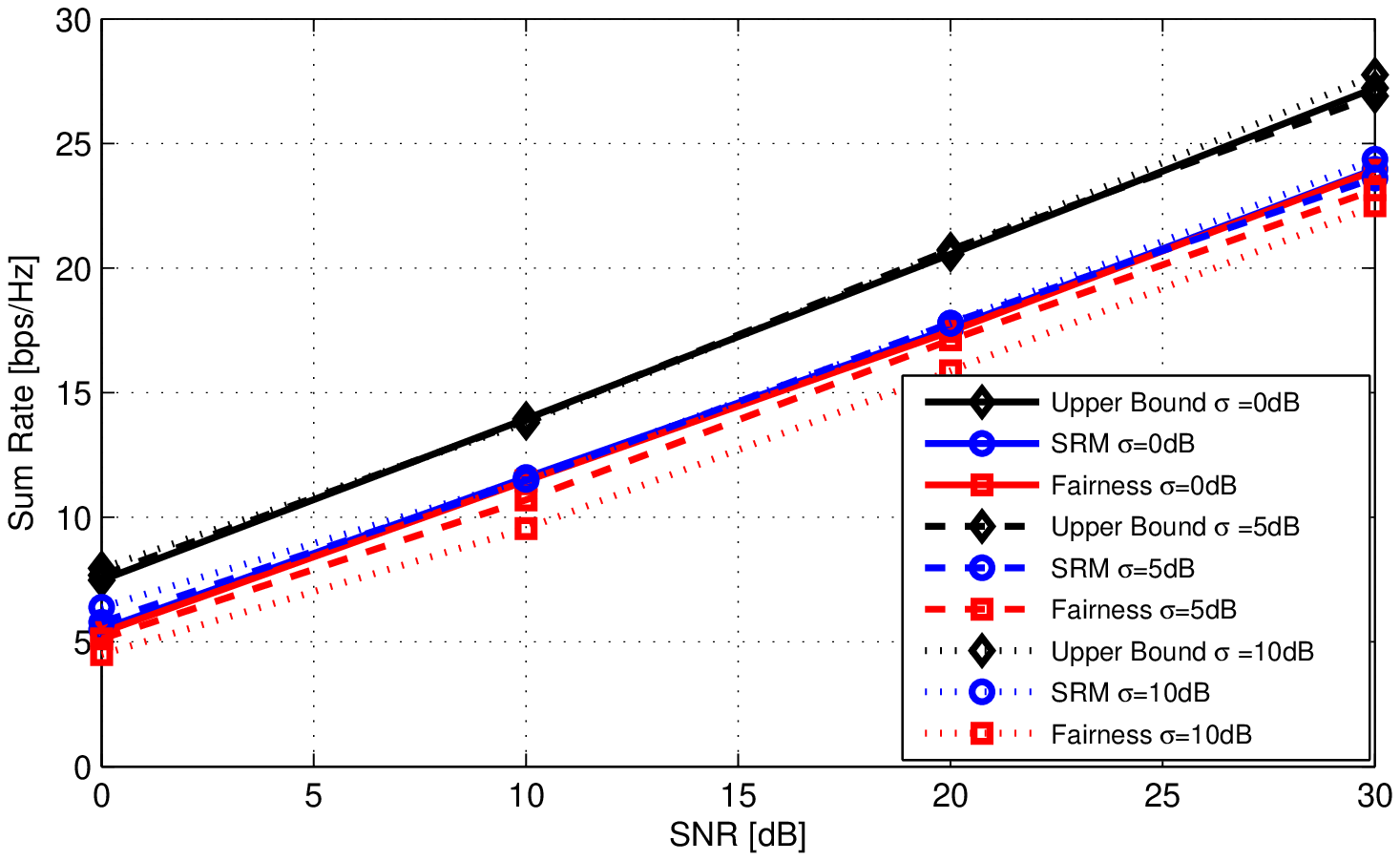}\label{fig:4.3b}}
\subfloat[]{\includegraphics[width=2.2in, height=1.7in]{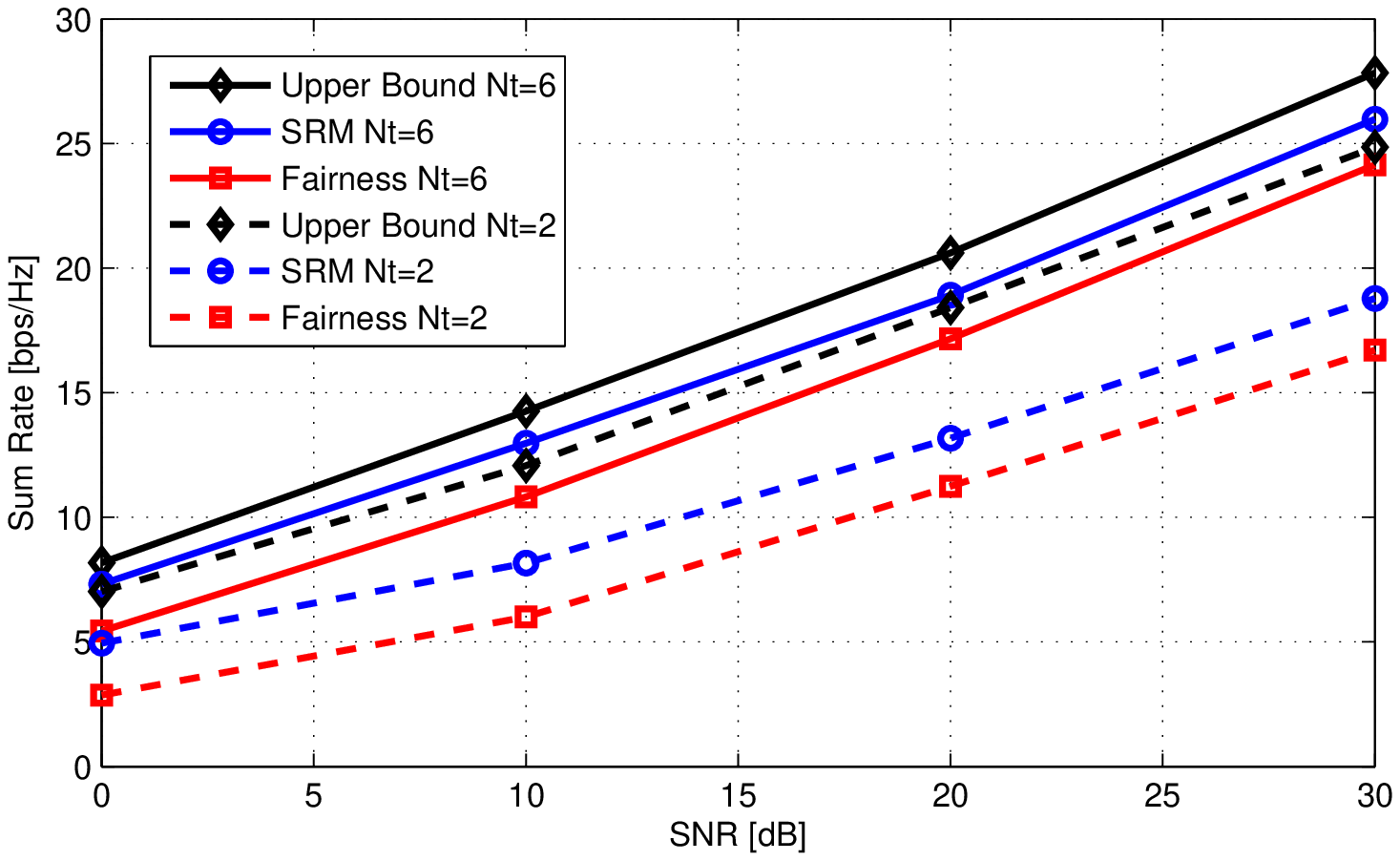}\label{fig:4.3a}}
\subfloat[]{\includegraphics[width=2.2in, height=1.7in]{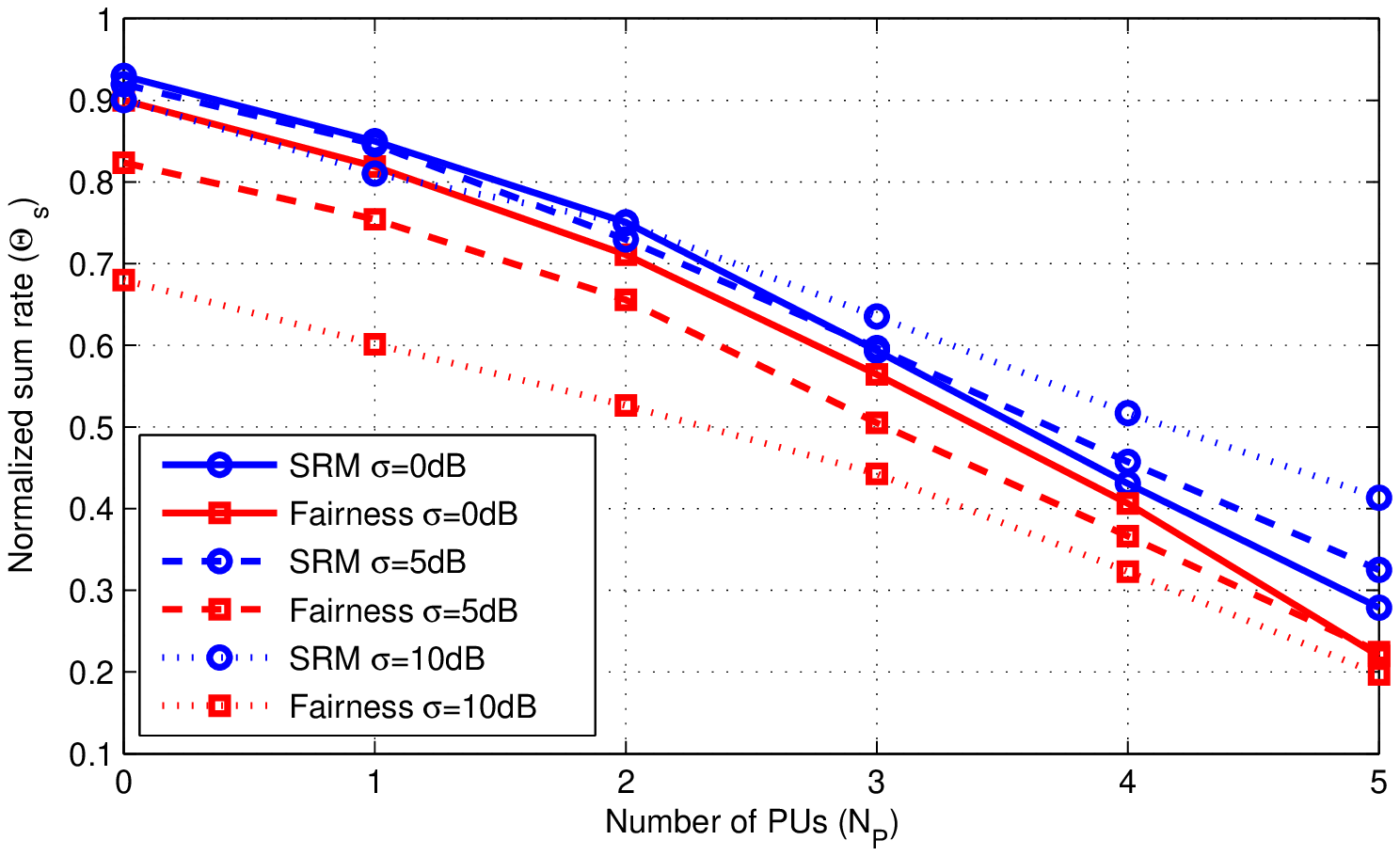}\label{fig:4.4}}
\caption{(a) Sum rate vs SNR for different SNR deviations ($N_t=6$ $N_r=4$, $N_s=2$, $N_p=2$), (b) Sum rate vs SNR for different antenna configurations ($N_r=4$, $N_s=2$, $N_p=1$, $\sigma=10dB$), (c) Normalized sum rate vs Number of PUs, for different SNR deviations ($N_t=6$ $N_r=4$, $N_s=4$, $SNR=10dB$)}
\label{fig:4.3}
\end{figure*}

This subsection evaluates the performance of both algorithms for the BC channel. The BC channel represents a specific multiuser scenario realization comprising one transmitter and multiple receivers. From the practical perspective the most frequent appearance of the BC channel is the downlink cell communication. \\
\indent Fig.~\ref{fig:4.3b} shows the achieved sum rate of both beamforming algorithms versus the SNR for different SNR deviations. As discussed at the beginning of this section, the SNR deviation is a valuable metric that can enable more comprehensive analysis for the behavior of the beamforming algorithms. Fig.~\ref{fig:4.3b} indicates that the SNR deviation does not impair the achieved sum rate of the SRM algorithm. However, it has significant impact on the performance of the Fairness algorithm, i.e. higher SNR deviations lead to decreased sum rates. Since the Fairness algorithm strives to balance the performance between the data streams, it will exploit the available resources on improving the rates of the data streams that experience lower SNR values and ultimately achieve lower sum rates compared to the SRM algorithm.       

\indent Fig.~\ref{fig:4.3a} depicts the achieved sum rate of both beamforming algorithms versus the SNR for different antenna configurations. Similar conclusions stand as for Fig.~\ref{fig:4.1}. The SRM algorithm always outperforms the Fairness algorithm for any SNR and antenna configuration. With the increasing the number of antennas in the system, both algorithms improve their performances and operate with rates that approach the upper bound. Fig.~\ref{fig:4.3a} also shows that the SRM performance gain (the sum rate difference) over the Fairness algorithm, is independent of the number of antennas. This is opposite to the conclusions deducted from the IC channel scenario. For the BC channel there exists a single/centralized transmission point that provides the possibility to allocate the available degrees of freedom and transmit power per user in a more efficient manner. This enables the SRM algorithm to \textit{opportunistically} adapt the transmit beamformer gain for each user in terms of the obtained channel conditions, and exploit the transmit power dimension in order to maximize the sum rate. \\
\indent Fig.~\ref{fig:4.4} depicts the normalized sum rate $\Theta _s$ of both algorithms in relation to the number of active PUs and different SNR deviations. As seen from the figure, both algorithms achieve lower sum rates for higher number of PUs. The increased number of PUs, stipulates the beamforming to exploit most of the available degrees of freedom to mitigate the inter PU-SU system interference and lead to decreased SU sum rates. However, for a small number of PUs both algorithms operate with rates that are very near the upper bound. With regards of the SNR deviation both algorithms exhibit contrasting behaviors. This is due to the reverse interference. For a small number of PUs the scenario is dominated primarily by the SNR and the same conclusions hold as for Fig.~\ref{fig:4.3b}. In contrast, for a large number of PUs the scenario is dominated by the reverse interference caused by the PU systems. Since the SRM algorithm predominantly maximizes the rates of the data streams with better channel conditions (in this case higher SINR), bigger SNR deviations will result in improved opportunities for the rate maximization process. With respect to the Fairness algorithm, the interference limited operation alleviates the SNR deviation effect on the algorithm's performance, as the majority of the SUs will exhibit unfavorable channels conditions (low SINRs) and result in convergence of the sum rates. Fig.~\ref{fig:4.4} also shows that both algorithms can be applied to a conventional BC channel beamforming scenario, i.e. BC channel without PU systems ($N_p=0$).      

\subsection{Multiple access channel}
\begin{figure*}[!t]
\centering
\subfloat[]{\includegraphics[width=2.2in, height=1.7in]{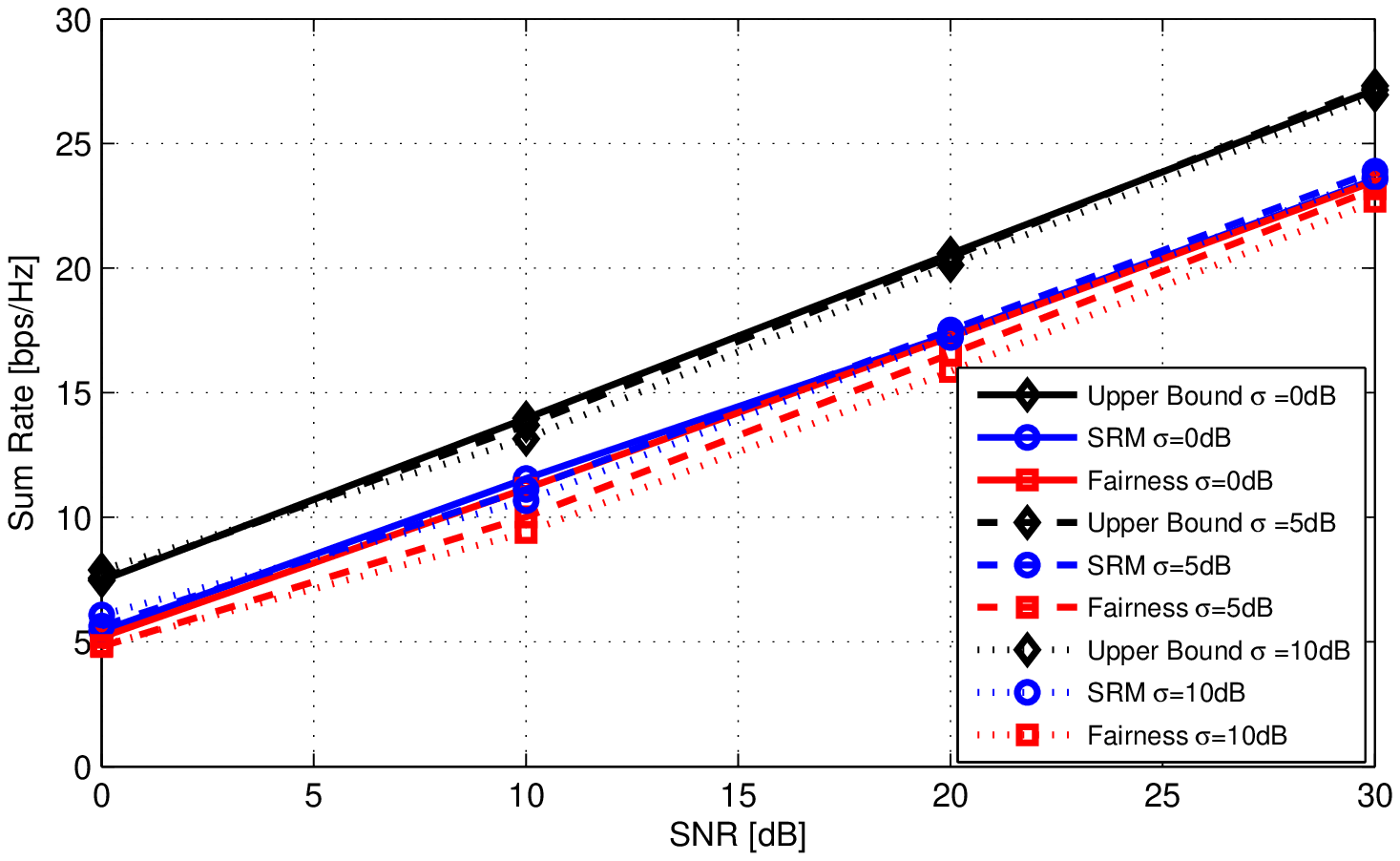}\label{fig:4.5a}}
\subfloat[]{\includegraphics[width=2.2in, height=1.7in]{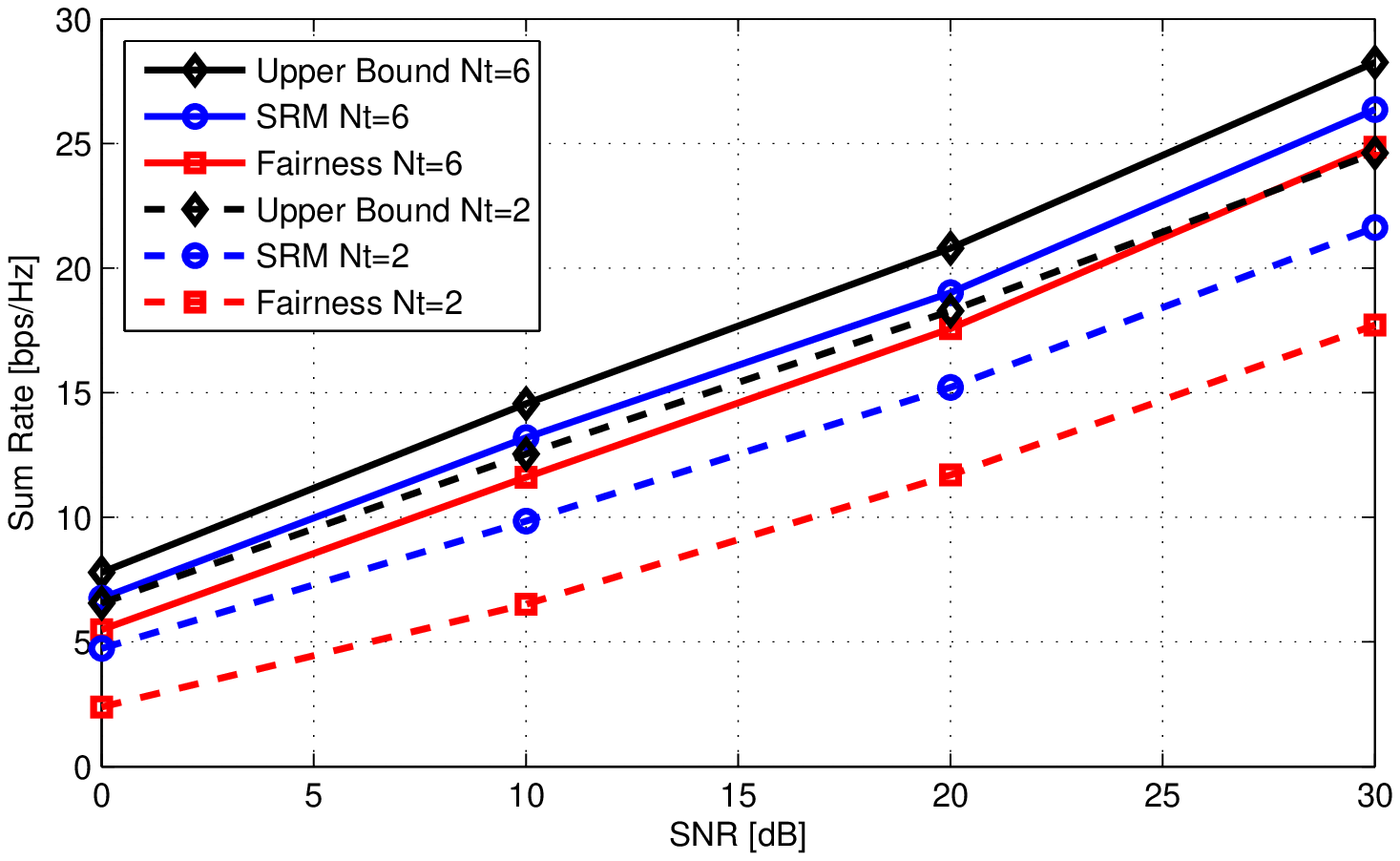}\label{fig:4.5b}}
\subfloat[]{\includegraphics[width=2.2in, height=1.7in]{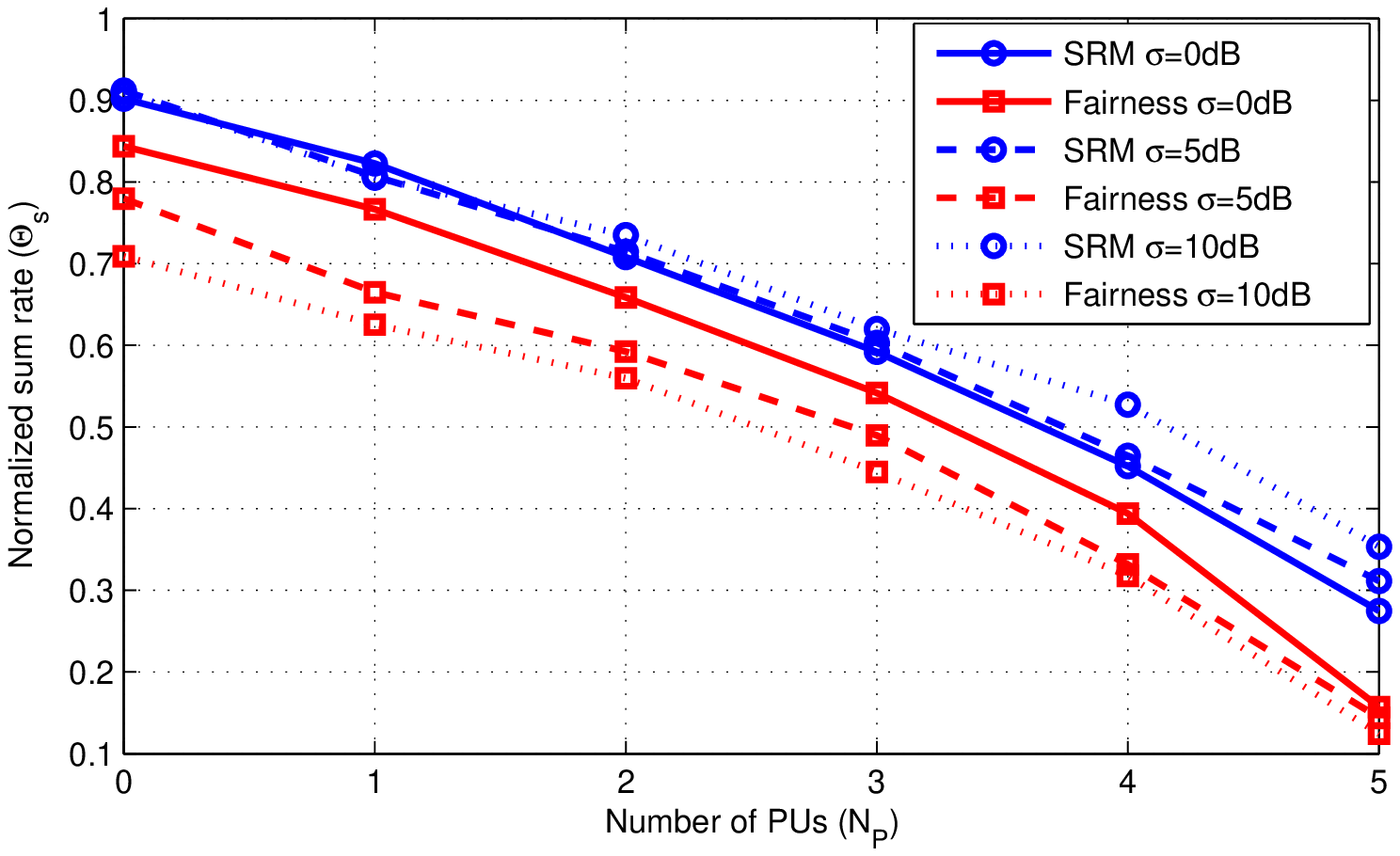}\label{fig:4.6}}
\caption{(a) Sum rate vs SNR for different SNR deviations ($N_t=6$ $N_r=4$, $N_s=2$, $N_p=2$), (b) Sum rate vs SNR for different antenna configurations ($N_r=4$, $N_s=2$, $N_p=1$, $\sigma=10dB$), (c) Normalized sum rate vs Number of PUs, for different SNR deviations ($N_t=6$ $N_r=4$, $N_s=4$, $SNR=10dB$)}
\label{fig:4.5}
\end{figure*}

In this subsection we consider the performance of both algorithms for the MAC channel. The MAC channel represents a specific multiuser scenario realization comprising multiple transmitters and one common receiver. From the practical perspective, the most frequent appearance of the MAC channel is the uplink of cell communication systems. \\
\indent Fig.~\ref{fig:4.5a} depicts the achieved sum rate of both beamforming algorithms versus the SNR and different SNR deviations. The same conclusions stand as for Fig.~\ref{fig:4.3b}. More specifically, the SNR deviation does not impair the achieved sum rate of the SRM algorithm, however, for the Fairness algorithm, higher SNR deviations lead to decreased sum rates.

Fig.~\ref{fig:4.5b} depicts the achieved sum rate of both beamforming algorithms versus the SNR and different antenna configurations. The SRM algorithm always outperforms the Fairness algorithm for any SNR and antenna configuration. With the increase of the number of antennas in the system, both algorithms improve their performances and operate with rates that approach the upper bound. Additionally, Fig.~\ref{fig:4.5b} shows that the SRM's performance gain over the Fairness algorithm diminishes for a higher number of antennas as a consequence of the distributed transmission system model (see, discussions for Fig.\ref{fig:4.1} and Fig.~\ref{fig:4.3a})\\
Fig.~\ref{fig:4.6} depicts the normalized sum rate $\Theta _s$ of both algorithms as a function of the number of active PUs and different SNR deviations. The same conclusions stand as for Fig.~\ref{fig:4.4}. Specifically, when the number of PUs is increased both algorithms achieve lower sum rates. Moreover, due to the interference limited operation of the SU system, higher SNR deviations improve the SRM algorithm's sum rate, but however, have almost no effect on the Fairness algorithm. For a small number of PUs both algorithms operate with rates that approach the upper bound. As a result of the SNR dominated operation of the SU system, higher SNR deviations decrease the performance of the Fairness algorithm, while having negligible impact on the SRM's performance. Fig.~\ref{fig:4.6} also shows that both algorithms can be applied to a conventional MAC channel beamforming scenario, i.e. MAC channel without PU systems ($N_p=0$).\\
\indent The presented results in Sec.IV.A, Sec.IV.B and Sec.IV.C, show that both algorithms are capable of efficient operation in all three system models (i.e. IC, BC and MAC), hence verifying the generic applicability of the proposed beamforming design (Sec.~\ref{sec:sysmodel}). Moreover, by comparing the results from Sec.IV.B and Sec.IV.C, it can be concluded that the SRM and the Fairness algorithms operate equally efficient in both downlink (BC channel) and uplink (MAC channel), demonstrating their universality. With respect to the achieved sum rates, the SRM algorithm always outperforms the Fairness algorithm, regardless of the system model and scenario setup. However, this is an anticipated result as the Fairness algorithm focuses on leveraging the rates between all data streams at the price of decreasing the overall sum rate. 

\subsection{Complexity reduction}
\begin{figure*}[!t]
\centering
\includegraphics[width=3.4in, height=2.1in]{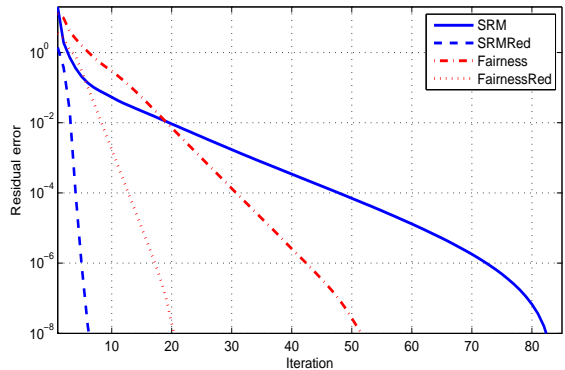}
\caption{Convergence rate comparison ($N_t=4$ $N_r=4$, $N_s=10$, $N_p=2$, $SNR=10dB$, $\sigma=5dB$)}
\label{fig:4.7}
\end{figure*}

This subsection assesses the convergence behavior of the SRM and the Fairness optimization, for the recursive and successive optimization cases. It verifies the advantages of the complexity reduction approach (Sec.III.C) in terms of the \textit{Convergence rate} of both algorithms. For the remainder of this subsection the successive SRM and Fairness optimization algorithms will be denoted as SRMRed and FairnessRed, respectively. Fig.~\ref{fig:4.7} depicts the Convergence rate of the SRM, Fairness, SRMRed and FairnessRed algorithms, for the IC channel scenario. The convergence rate is presented as the residual error (the difference between the attained value and the optimal value\footnote[1]{The values of interest represent the optimization functions values for the SMR (\ref{eq:29}) and the Fairness (\ref{eq:18}) optimization, i.e. the sumrate and the fairness coefficient $\tau$, respectively.}) in function of the number of iterations performed by the beamforming algorithm. The IC channel is specifically chosen for the performance evaluation in Fig.~\ref{fig:4.7}, as the most complex of the three system models, and most suitable for emphasizing the convergence behavior of the beamforming algorithms.    

Fig.~\ref{fig:4.7} shows that the SRM algorithm requires a higher number of iterations in order to converge to the optimal solution when compared to the Fairness algorithm. It is also evident that the complexity reduction approach significantly improves the convergence rates of both beamforming algorithms, and can facilitate their operation in scenarios which require agile beamformer computation. Additionally, Fig.~\ref{fig:4.7} shows that the SRMRed algorithm achieves higher convergence rate in comparison to the FairnessRed algorithm. \\
This section evaluated the performances of the SRM and the Fairness algorithms. It verified their universal applicability and rate efficiency for for variety of parameters and scenario setups. Moreover, it validated their underlying complexity and possible reduction by applying the approach presented in Sec.III.C.            

\section{Conclusion}
\label{sec:Concl}
\indent Recently, the concept of MIMO and beamforming has been extensively investigated as a possible enabler of the underlay spectrum sharing. A common limitation of earlier proposed beamforming techniques is that they either cannot exploit arbitrary number of antennas in the system and/or cannot serve an arbitrary number of secondary data streams in the system. Moreover, the existing beamforming techniques frequently treat the reverse interference as noise at the secondary receiver. Additionally, all of the existing beamforming works we are aware of that focus on underlay spectrum sharing propose algorithms that lack universality and are specifically designed either for the interference channel, broadcast channel or the multiple access channel. This segregated design significantly limits their applicability in practical scenarios, where the underlying beamforming process should be capable of operating in any possible multiuser scenario. \\
\indent In this paper we have proposed several advancements and novel techniques with respect to the beamforming based underlay spectrum sharing. First, the paper extends the commonly utilized underlay spectrum sharing model by incorporating the reverse interference in the beamforming process. Additionally it proposes a generic combined beamforming design that is applicable for any multiuser scenario and is applicable for conventional as well as for underlay spectrum sharing based systems. Furthermore, the paper develops two novel multiuser beamforming recursive algorithms for user fairness and sum rate maximization based on the proposed generic beamforming design. Both algorithms provide operate on convex problems for the computation of the optimal transmit and receive beamformers. The paper also elaborates on a possible successive optimization approach that decreases the complexity of the proposed fairness and sum rate maximization beamforming algorithms. The presented results in the paper clearly show that both beamforming algorithms are capable of efficient operation in any multiuser scenario. Both algorithms can operate for any number of antennas and users in the system, thus verifying their universality and practical applicability. Moreover, the results show that the proposed complexity reduction approach significantly improves the convergence rates of both algorithms and can facilitate their operation in scenarios which require agile beamformer computation. 

\section*{Acknowledgment}
The work was sponsored by the Public Diplomacy Division of NATO in the framework of “Science for Peace” through the SfP-984409 “Optimization and Rational Use of Wireless Communication Bands (ORCA)” project and inspired discussions and work done in the context of European Union supported Network of Excellence project ACROPOLIS (FP7-257626). One of us (PM) acknowledges also funding from DFG (Deutsche Forschungsgemeinschaft). 

\ifCLASSOPTIONcaptionsoff
  \newpage
\fi

\bibliographystyle{IEEEtranTCOM}
\bibliography{IEEE_TCOM}
\end{document}